\documentclass[desactivate]{aa}
 
\bibpunct{(}{)}{;}{a}{}{,} % to follow the A&A style

\usepackage{graphicx}
\usepackage[varg]{txfonts}
\usepackage{hyperref}
\usepackage{orcidlink}
\usepackage{amsmath}
\usepackage{xcolor}
\usepackage[normalem]{ulem}
\usepackage{booktabs}

\hypersetup{
  colorlinks=true,
  allcolors=blue
}

\begin{document}

\title{Hanging on the cliff: Extreme mass ratio inspiral formation with local two-body relaxation and post-Newtonian dynamics}
\titlerunning{Hanging on the cliff: EMRI formation with local two-body relaxation and PN dynamics}

\author
{
    Davide Mancieri
    \inst{1}\fnmsep\inst{2}\fnmsep\inst{3}\fnmsep
    \thanks{\email{d.mancieri@campus.unimib.it}}
    \!\orcidlink{0009-0004-5106-9363}
    \and
    Luca Broggi
    \inst{2}\fnmsep\inst{3}
    \!\orcidlink{0000-0002-9076-1094}
    \and
    Matteo Bonetti
    \inst{2}\fnmsep\inst{3}\fnmsep\inst{4}
    \!\orcidlink{0000-0001-7889-6810}
    \and
    Alberto Sesana
    \inst{2}\fnmsep\inst{3}\fnmsep\inst{4}
    \!\orcidlink{0000-0003-4961-1606}
}
\authorrunning{Davide Mancieri et al.}

\institute
{
    Dipartimento di Fisica, Università degli Studi di Trento,
    Via Sommarive 14, 38123 Povo, Italy
    \and
    Dipartimento di Fisica ``G. Occhialini'', Università degli Studi di Milano-Bicocca,
    Piazza della Scienza 3, 20126 Milano, Italy
    \and
    INFN - Sezione di Milano-Bicocca,
    Piazza della Scienza 3, 20126 Milano, Italy
    \and
    INAF - Osservatorio Astronomico di Brera,
    Via Brera 20, 20121 Milano, Italy
}

\date{Received 19 September 2024 / Accepted 20 January 2025}

\abstract
{    
    Extreme mass ratio inspirals (EMRIs) are anticipated to be primary gravitational wave sources  for the Laser Interferometer Space Antenna (LISA). They form in dense nuclear clusters when a compact object is captured by the central massive black holes (MBHs) as a consequence of the frequent two-body interactions occurring between orbiting objects. The physics of this process is complex and requires detailed statistical modelling  of a multi-body relativistic system. 
    We present a novel Monte Carlo approach to evolving the post-Newtonian (PN) equations of motion of a compact object orbiting an MBH. The approach accounts for the effects of two-body relaxation locally on the fly, without leveraging on the common approximation of orbit-averaging. We applied our method to study the function $S(a_0)$, describing the fraction of EMRI to total captures (including EMRIs and direct plunges, DPs) as a function of the initial semi-major axis $a_0$ for compact objects orbiting central MBHs with $M_\bullet\in[10^4 \,{\rm M}_\sun, 4\times10^6 \,{\rm M}_\sun]$. The past two decades have consolidated a picture in which $S(a_0)\rightarrow 0$ at large initial semi-major axes, with a sharp transition from EMRIs to DPs occurring around a critical scale $a_{\rm c}$. A recent study challenges this notion for  low-mass MBHs, finding EMRIs forming at $a\gg a_{\rm c}$, which were called `cliffhangers'. Our simulations confirm the existence of cliffhanger EMRIs, which we find to be more common then previously inferred. Cliffhangers start to appear for $M_\bullet\lesssim3\times 10^5 \,{\rm M}_\sun$ and can account for up to 55\% of the overall EMRIs forming at those masses. We find $S(a_0) \gg 0$ for $a \gg a_{\rm c}$, reaching values as high as 0.6 for $M_\bullet = 10^4 \,{\rm M}_\sun$, much higher than previously found. We test how these results are influenced by different assumptions on the dynamics used to evolve the system and treatment of two-body relaxation. We find that the PN description of the system greatly enhances the number of EMRIs by shifting $a_{\rm c}$ to larger values at all MBH masses. Conversely, the local treatment of relaxation has a mass-dependent impact, significantly boosting the number of cliffhangers at low MBH masses compared to an orbit-averaged treatment.
    These findings highlight the shortcomings of standard approximations used in the EMRI literature and the importance of carefully modelling the (relativistic) dynamics of these systems. The emerging picture is more complex than previously thought, and should be considered in future estimates of rates and properties of EMRIs detectable by LISA.

}


\maketitle

%%%%%%%%%%%%%%%%%%%%%%%%%%%%%%%%%%%%%%%%%%%%%%%%%%%%%%%%%%%%%%%%%%%%%%%%%%%%
%%%%%%%%%%%%%%%%%%%%%%%%%%%%%%%%%%%%%%%%%%%%%%%%%%%%%%%%%%%%%%%%%%%%%%%%%%%%

\section{Introduction}

   Most galaxies host a massive black hole (MBH) at their core, with masses in the range of $10^5 \text{-} 10^{10} \, \mathrm{M}_\sun$ \citep{1995ARA&A..33..581K,1998AJ....115.2285M,2013ARA&A..51..511K}. Often these MBHs are found at the centre of nuclear clusters \citep{1997AJ....114.2366C,2002AJ....123.1389B,2006ApJS..165...57C,2007A&A...469..125S,2009A&A...502...91S,2012ApJS..203....5T}: highly dense clusters of stars and compact objects (COs) observed in the nuclei of many galaxies. In the central region of nuclear clusters, densities can reach up to $10^7 \, \mathrm{M}_\sun  \mathrm{pc}^{-3}$ \citep{2020A&ARv..28....4N}, facilitating the scattering of COs onto tight and eccentric orbits around the central MBH, thus forming  binaries with extreme mass ratios \citep{2008gady.book.....B,2013degn.book.....M}.
   
   General relativity (GR) predicts that these binary systems emit gravitational waves (GWs), mostly during each CO pericentre passage \citep{1973grav.book.....M}. Over time the CO orbit shrinks and circularises \citep{1964PhRv..136.1224P} as a consequence of the loss of energy and angular momentum of the binary system, which are carried away by GWs \citep{10.1093/acprof:oso/9780198570745.001.0001}. Accordingly, the CO slowly inspirals  towards the MBH, completing thousands of orbital cycles, eventually merging with it \citep{2018LRR....21....4A}. These GW sources are known as extreme mass ratio inspirals (EMRIs) due to the large mass imbalance between the pair of objects that make up the binary system: the MBH and the orbiting CO \citep{2007CQGra..24R.113A,2022hgwa.bookE..17A}. Alternatively, if the merger occurs after few orbits or in a head-on collision, the event is referred to as a direct plunge (DP).

   Extreme mass ratio inspirals that form around MBHs with masses  of the order of $10^4 \text{-} 10^7 \, \mathrm{M}_\sun$ are expected to emit GWs in the millihertz frequency range \citep{2004PhRvD..69h2005B,2018LRR....21....4A}. The numerous bursts of GWs emitted during an EMRI are anticipated to build up a significant signal-to-noise ratio \citep{2004PhRvD..69h2005B,2017PhRvD..95j3012B}, enabling the detection of such events by the upcoming Laser Interferometer Space Antenna \citep[LISA,][]{2017arXiv170200786A,2023LRR....26....2A,2024arXiv240207571C}. Detecting EMRIs will enable us to test GR \citep{1997PhRvD..56.1845R,2007PhRvD..75d2003B,2024arXiv240602454G}, estimate cosmological parameters \citep{2008PhRvD..77d3512M,2021MNRAS.508.4512L}, and investigate dormant MBHs \citep{2010PhRvD..81j4014G,2019ApJ...883L..18G}.
   
   Recent works \citep[e.g.][]{2022ApJ...926..101M,2023A&A...675A.100F,2023ApJ...957...34L} have also proposed that EMRIs might be responsible for the novel, currently unexplained, X-ray variability phenomenon of quasi-periodic eruptions \citep{2019Natur.573..381M,2020A&A...636L...2G,2021Natur.592..704A,2024A&A...684A..64A}, and recent observations agree with such an interpretation \citep{2024Natur.634..804N}. This would offer a way to detect and study EMRIs within the electromagnetic spectrum, enabling multi-messenger observations that would boost their potential as astrophysical and cosmological probes \citep[e.g.][]{2021PhRvL.126j1105T}.
   
   Extreme mass ratio inspirals can form in many different ways, including (i) the tidal separation of a stellar-mass black hole (BH) binary system by the central MBH followed by the capture of one of the binary components \citep{1988Natur.331..687H,2005ApJ...631L.117M}; (ii) the formation or capture and subsequent migration of BHs within accretion discs surrounding MBHs \citep{2007MNRAS.374..515L,2021PhRvD.103j3018P}; (iii) the capture of a BH remnant by an MBH as a consequence of supernova kicks \citep{2017MNRAS.469.1510B,2019MNRAS.485.2125B}; and (iv) the scatter of BHs onto tight orbits due to the gravitational influence of a secondary MBH in the aftermath of galaxy mergers \citep{2011ApJ...729...13C,2014MNRAS.438..573B,2022ApJ...927L..18N,2022MNRAS.516.1959M}.
   
   The standard and ubiquitous formation channel for EMRI formation, which we explore in this work, is the scatter of a CO onto a sufficiently tight and eccentric orbit as a consequence of two-body relaxation \citep{2017ARA&A..55...17A,2018LRR....21....4A}. The estimated EMRI formation rate through this channel is in the range of $10^{-8} \text{-} 10^{-6}$ EMRIs per year for a central MBH of mass $M_\bullet = 10^6 \, \mathrm{M}_\sun$ \citep{1995ApJ...445L...7H,2004CQGra..21S1595G,2006ApJ...649L..25R}. Considering also the uncertainty in the number density of MBHs inside the observable range of LISA and the fraction of these that are hosted in nuclear clusters, the predicted number of EMRIs that will be detected by LISA during its lifetime spans   three orders of magnitude, from a few events per year to a few thousand \citep{2017PhRvD..95j3012B}.

   Simulating the formation of EMRIs in this standard channel poses considerable challenges. The most accurate description of the whole nuclear cluster would involve a direct N-body computation of the orbits of all stars and COs within it. However, the need to follow these objects for thousands of orbits, and the high mass ratio between them and the central MBH pose an insurmountable challenge for N-body approaches, which show problems even in the simplest Newtonian regime \citep{2018LRR....21....4A}. Alternative methods, such as solving the steady-state Fokker-Planck equation for the distribution function of the system \citep[e.g.][]{2011CQGra..28i4017A,2016ApJ...820..129B,2017ApJ...848...10V,2022MNRAS.514.3270B,2024arXiv240607627K} or employing Monte Carlo codes \citep[e.g.][]{1971Ap&SS..13..284H,1998MNRAS.298.1239G,2000ApJ...540..969J,2001A&A...375..711F,2013ApJS..204...15P,2018ApJ...852...51F,zhangMonteCarloStellarDynamics2023a}, offer faster approximated solutions, but carry inherent limitations and assumptions. 
   
   From the aforementioned body of work a picture has emerged in which there exists a critical initial semi-major axis $a_\mathrm{c}$ for the CO orbit around the MBH determining its eventual fate \citep{2005ApJ...629..362H,2016ApJ...820..129B,2021MNRAS.501.5012R}. In short, COs starting their inspiral from orbits with $a_0 < a_\mathrm{c}$ would only result in the formation of EMRIs, while only DPs would form when $a_0 > a_\mathrm{c}$, with a sharp transition between the two regimes around $a_\mathrm{c}$. Notably, this finding has been derived from detailed simulations of nuclear clusters hosting an MBH with $M_\bullet \geq 10^6\, {\rm M}_\odot$ and then extrapolated at lower masses. 
   Recently, however, this dichotomy has been questioned in \citet{2024PhRvL.133n1401Q} (hereafter QS24) for the case of  low-mass MBHs and intermediate-mass BHs, opening to the possibility of EMRIs forming even from initially wide orbits, dubbed  cliffhanger EMRIs. 
   This is an intriguing claim that might bear interesting consequences for estimating LISA EMRI detection rates. 

   As mentioned above, the techniques and codes employed to verify both the existence of $a_\mathrm{c}$ and the appearance of cliffhanger EMRIs rely on a number of assumptions and approximations. 
   A notable approximation is that of being allowed to average the effects on the orbit of two-body relaxation over an entire orbital period \citep[e.g.][QS24]{1978ApJ...226.1087C,2005ApJ...629..362H}. However, this simplification might not accurately describe the system, as two-body encounters strongly depend on the instantaneous velocity of the orbiting CO \citep{2008gady.book.....B,2013degn.book.....M}, which significantly changes during a highly eccentric orbit. Another  is to use Newtonian dynamics to describe the system. GR effects are usually included only by adding extra orbit-averaged loss terms due to GW emission and by adjusting the CO capture condition by a suitable factor. 
    
   %In this work we present a novel post-Newtonian (PN) Monte Carlo code to locally treat two-body relaxation multiple times within a single orbit. We then apply this code to study the fraction of EMRIs relative to DPs that form around an MBH, given an initial semi-major axis of the orbit $a_0$.
   
 In this work we present a novel post-Newtonian (PN) Monte Carlo code that evolves the PN dynamics of a CO orbiting an MBH. The code treats two-body relaxation locally, by applying multiple small kicks to the CO orbit within a single orbit, and can account for an arbitrary number of stellar components contributing to relaxation and to the potential in which the CO is evolving. As a first application, here we apply this code to study the fraction of EMRIs relative to DPs that form around MBHs of different masses as a function of the initial semi-major axis of the orbit $a_0$. This allows us to verify the existence of $a_\mathrm{c}$ and its dependence on the system parameters, and the emergence of the novel population of cliffhanger EMRIs found in QS24.
      
   %One of these assumptions is that of being allowed to average the effects on the orbit of two-body relaxation over an entire orbital period \citep[e.g.][]{1978ApJ...226.1087C,2005ApJ...629..362H,2023arXiv230413062Q}. However, such simplification might not accurately describe the system, as two-body encounters strongly depend on the instantaneous velocity of the orbiting CO \citep{2008gady.book.....B,2013degn.book.....M}, which significantly changes during a highly eccentric orbit.

   %To address this question, in this work we present a post-Newtonian (PN) Monte Carlo code to locally treat two-body relaxation multiple times within a single orbit. We then apply this code to study the fraction of EMRIs relative to DPs that form around an MBH, given an initial semi-major axis of the orbit $a_0$.

   %Previous works \citep{2005ApJ...629..362H,2016ApJ...820..129B,2021MNRAS.501.5012R} have identified a critical value of the semi-major axis $a_\mathrm{c}$ such that COs starting their inspiral from orbits with $a_0 < a_\mathrm{c}$ should only result in the formation of EMRIs, while no EMRI should occur for $a_0 > a_\mathrm{c}$. Recently, however, such dichotomy has been questioned in \citet{2023arXiv230413062Q} (hereinafter QS24) for the case of low-mass MBHs and intermediate-mass BHs, opening to the possibility of EMRIs forming even from initially wide orbits, dubbed \textit{cliffhanger EMRIs}. Verifying such claim is of fundamental importance in order to accurately estimate LISA detection rates and model the growth of MBHs.

   The study presented in QS24 has some limitations which our code can improve upon. It employs a Monte Carlo approach to evolve in time the Newtonian energy and angular momentum of an orbit subjected to two-body relaxation, which is orbit-averaged over more than a full orbit at times. The loss of energy and angular momentum due to GW radiation is accounted for through equations detailed in \citet{1964PhRv..136.1224P}, which are orbit-averaged as well. Moreover, only two-body encounters between a stellar-mass BH of mass $m_\bullet = 10 \, \mathrm{M}_\sun$ and a single-mass population of $m_\star = 1 \, \mathrm{M}_\sun$ stars surrounding the central MBH are considered. The stars are assumed to follow a simple isotropic power-law numeric density, and their global gravitational potential is not considered. In comparison, in this work we track the orbits of a stellar-mass BH in the total potential well of an MBH and of two populations of objects surrounding it: stars with masses $m_\star$ and stellar-mass BHs with masses $m_\bullet$. We describe both populations with Dehnen numeric density profiles \citep{1993MNRAS.265..250D,1994AJ....107..634T}. We also solve the equations of motion up to the 2.5PN term,\footnote{In what follows, we call `$n$PN term' the term of the Hamiltonian defined in Eq.\ \eqref{eq:hamiltonian}, which is suppressed by a factor $(1/c)^{2n}$ with respect to the leading term.} thus naturally accounting for GW radiation, and we locally correct the shape of the orbit for two-body relaxation multiple times within each orbital period.

   %The study presented in QS24 has some limitations which we lift. It employs a Monte Carlo approach to evolve in time the Newtonian energy and angular momentum of an orbit due to two-body relaxation, where such effects are orbit-averaged over more than a full orbit at times. The loss of energy and angular momentum due to GW radiation is accounted for through equations detailed in \citet{1964PhRv..136.1224P}, which are orbit-averaged as well. Moreover, only two-body encounters between a stellar-mass BH of mass $m_\bullet = 10 \, \mathrm{M}_\sun$ and a single-mass population of $m_\star = 1 \, \mathrm{M}_\sun$ stars surrounding the central MBH are considered. The stars are assumed to follow a simple isotropic power-law numeric density, and their global gravitational potential is not considered.   
   
   %In comparison, in this work we track the orbiting of a stellar-mass BH in the total potential well of an MBH and of two populations of objects surrounding it: stars with masses $m_\star$ and stellar-mass BHs with masses $m_\bullet$. We describe both populations with Dehnen numeric density profiles \citep{1993MNRAS.265..250D,1994AJ....107..634T}. We also solve the equations of motion up to the 2.5PN term\footnote{In what follows, we call ``$n$PN term'' the term of the Hamiltonian defined in Eq.\ \eqref{eq:hamiltonian} which is suppressed by a factor $(1/c)^{2n}$ with respect to the leading one.}, thus naturally accounting for GW radiation, and locally correct the shape of the orbit for two-body relaxation multiple times within each orbital period.

   The paper is organised as follows. In Sect.\ \ref{sec:theory} we review the aspects of the theory relevant to our implementation and predict some behaviours of cliffhanger EMRIs with analytical means. In Sect.\ \ref{sec:methods} we delve into the specifics of our methods, to then present the results of our simulations in Sect.\ \ref{sec:results}. Finally, in Sect.\ \ref{sec:disc_conc} we discuss our findings and draw our conclusions.

%__________________________________________________________________

\section{Theory of relaxation-driven EMRI formation} \label{sec:theory}   

    In this section we first introduce some notions on the treatment of two-body interactions in the loss cone framework. Then, we give operative definitions for EMRIs and DPs. Finally, we briefly review the literature on the study of the EMRI-to-DP ratio and draw some first theoretical results on cliffhanger EMRIs.

    \subsection{Loss cone and diffusion theory} \label{sec:losscone_diff}

        The main actor in a nuclear cluster is the central MBH, as it strongly affects the motion of stars and COs within its influence radius, which we define in this work as \citep{1972ApJ...178..371P}
        \begin{equation}
            R_\mathrm{inf} = \frac{G M_\bullet}{\sigma^2} \, .
        \end{equation}
        Here $G$ is the gravitational constant, $M_\bullet$ is the mass of the MBH, and $\sigma$ is the velocity dispersion of stars in the bulge of the host galaxy. The velocity dispersion can be related to the mass of the central MBH through the M--$\sigma$ relation \citep{2000ApJ...539L...9F,2000ApJ...539L..13G}, that we employ using the best fit from \citet{2009ApJ...698..198G}:
        \begin{equation}
            \sigma = 70 \left( \frac{M_\bullet}{1.53 \times 10^6 \, \mathrm{M}_\sun}\right)^{1/4.24} \, \mathrm{km/s} \, .
        \end{equation}
The nuclear star cluster is then composed by many other celestial bodies, which can be grouped into populations based on their nature and mass (e.g. 
Sun-like stars, stellar-mass BHs, neutron stars, and so on). We start by considering an arbitrary number of populations, later specialising to a system with only two (which can be identified generically as solar-mass stars and stellar-mass BHs).
        
        We label each population with an $\alpha$ index\footnote{To give a concrete example, we use $\alpha = \star$ to refer to a population of solar-mass stars, and $\alpha = \bullet$ to refer to a population of stellar-mass BHs.} and model their number distribution with Dehnen profiles \citep{1993MNRAS.265..250D,1994AJ....107..634T}:
        \begin{equation} \label{eq:n}
            n_\alpha (r) = \frac{3-\gamma_\alpha}{4\pi}\frac{M_\mathrm{tot,\alpha}}{m_\alpha}\frac{r_\mathrm{a,\alpha}}{r^{\gamma_\alpha} \, (r+r_\mathrm{a,\alpha})^{4-\gamma_\alpha}} \, .
        \end{equation}
        Here $m_\alpha$ is the individual mass of the objects in the $\alpha$ population, while $M_\mathrm{tot,\alpha}$ is their total mass. The exponent $\gamma_\alpha$, comprised in the range $[0,3)$, is the slope of the profile for $r \to 0$ and $r_\mathrm{a,\alpha}$ is the scale radius at which the distribution smoothly switches to a slope of -4 for $r \to \infty$. Finally, $r$ measures the distance from the centre of the distribution. We consider each profile in a steady-state configuration, with their centres fixed at the origin of our coordinate system, just like the MBH.
        %\as{Question (pardon my ignorance): Dehnen profiles are equilibrium solutions of Poisson equation. Is a superposition of Dehnen profiles with different scale radius also an equilibrium solution because of linearity?}\lb{The thing is not linear exactly: if you set the mass density to a sum of two Dehnen profiles (as we do) the potential changes linearly. However, when you compute the distribution function it is not the sum of two Dehnens, as the map $(r,v)\to E$ changes due to the different potential. However, you can still find an isotropic distribution for each component, therefore of the Jeans form, and it is a solution of the collisionless problem (so in the limit of a number of particles $N\to\infty$ where scatterings are negligible, it will stay there forever).}

        The gravitational potential produced by population $\alpha$ is \citep{1993MNRAS.265..250D,1994AJ....107..634T}
        \begin{equation} \label{eq:phi}
            \phi_\alpha (r) =
                \begin{cases}
                    \displaystyle - \frac{1}{2-\gamma_\alpha} \left(1- \left( \frac{r}{r+r_\mathrm{a,\alpha}}\right)^{2-\gamma_\alpha}\right) \frac{GM_\mathrm{tot,\alpha}}{r_\mathrm{a,\alpha}} & \text{if } \gamma_\alpha \neq 2 \\
                    \displaystyle \ln{\left(\frac{r}{r+r_\mathrm{a,\alpha}}\right)} \frac{GM_\mathrm{tot,\alpha}}{r_\mathrm{a,\alpha}} & \text{if } \gamma_\alpha = 2
                \end{cases} \, ,
        \end{equation}
        so the total potential at a given position is given by
        \begin{equation} \label{eq:psi}
            \psi (r) = - \, \left( - \frac{GM_\bullet}{r} + \sum_\alpha \phi_\alpha (r) \right) \, ,
        \end{equation}
        where we flipped the overall sign adopting the convention that the potential experienced by any object in the distribution is positive definite.

        The central MBH affects the distribution of surrounding objects also by removing bodies that pass too close to it at orbital pericentre. For example, stars can be tidally disrupted and subsequently accreted or directly swallowed, whereas COs can be gravitationally captured resulting in EMRIs and DPs. 
        This happens if the velocity vector $\vec{v}$ of a bound object points towards the central MBH within a small solid angle, inside the so-called loss cone: a cone-shape region in velocity space containing objects which will be lost inside the MBH at the pericentre of their current orbit \citep[see][]{2017ARA&A..55...17A,2018LRR....21....4A}. In nuclear clusters, objects can be brought closer to or directly inside the loss cone as a consequence of two-body encounters within each other. 

        Let us now consider a particular stellar-mass BH of mass $m_\bullet = 10 \, \mathrm{M_\sun}$ and velocity magnitude $v$ having a sequence of two-body encounters with the background bodies; we decompose its velocity change as the sum of $\Delta v_\parallel$, parallel to the initial line of motion, and $\Delta v_\perp$, in the plane orthogonal to the original direction of motion. At any point, given the position $r$ and velocity $v$ of the stellar-mass BH, its orbital binding energy per unit mass can be computed as
        \begin{equation}
            \mathcal{E} (r,v) = - \left( \frac{H (r,v)}{m_\bullet} + \sum_\alpha \phi_\alpha (r) \right) \, ,
        \end{equation}
        where $H$ is the Hamiltonian of the binary comprised of the MBH and the orbiting stellar-mass BH, and we define the energy so that $\mathcal{E} < 0$ for bound orbits. We report the expression of $H$ we used in this work in Eq.\ \eqref{eq:hamiltonian}.
        
        Given the random nature of the scattering process, different encounters will result in different values for $\Delta v_\parallel$ and $\Delta v_\perp$. It is for this reason that the problem is better approached statistically, by looking at the diffusion coefficients (i.e.\ the expectation values per unit of time) of $\Delta v_\parallel$, $\Delta v_\perp$, and related quantities. We use the symbol $\mathrm{D}[X]$ for the diffusion coefficient of the generic quantity $X$, and we explicitly write those relevant to our study in Appendix \ref{app:diff_coe}. Crucially, these are local diffusion coefficients, meaning they depend on the exact position and velocity of the orbiting stellar-mass BH. The usual approach in literature is instead that of employing orbit-averaged diffusion coefficients $\mathrm{D_o}[X]$ to account for two-body encounters, which follow from their local versions as \citep{2008gady.book.....B,2013degn.book.....M}:
        \begin{equation}
            \mathrm{D_o}[X] = \frac{2}{P} \int^{r_+}_{r_-} \frac{\mathrm{d}r}{v_\mathrm{r}} \mathrm{D}[X] \, .
        \end{equation}
        Here $r_+$ and $r_-$ respectively indicate the apocentre and pericentre of the orbit, $P$ its radial period, and $v_\mathrm{r}$ the radial component of the velocity. If we indicate with $a$ and $e$ the semi-major axis and the eccentricity of an orbit, their Keplerian estimates are
        \begin{equation} \label{eq:peri_apo}
            r_+ = a(1+e) \, , \qquad r_- = a(1-e) \, 
        \end{equation}
        and
        \begin{equation} \label{eq:period}
            P = 2 \pi \left(\frac{a^3}{G (M_\bullet + m_\bullet)} \right)^{1/2} \, .
        \end{equation}
        We note that the orbit is an ellipse with semi-major axis $a$ and eccentricity $e$ only in the Keplerian approximation, that holds where the stellar potential is negligible ($r_+ \lesssim R_\mathrm{inf}$), and the orbit is non-relativistic ($r_-\gg R_\mathrm{g}$). Here
        \begin{equation}
            R_\mathrm{g} = \frac{GM_\bullet}{c^2}
        \end{equation}
        is the gravitational radius of the MBH, where $c$ is the speed of light in vacuum. For orbits where the Keplerian approximation does not hold, our definitions of $a$ and $e$ do not have a simple, direct interpretation (see Sect.\ \ref{sec:EMRI/DP} below).
        
        In Appendix \ref{app:diff_coe} we show that a key ingredient to compute the local diffusion coefficients is $f_\alpha (\mathcal{E})$: the ergodic distribution function of objects in population $\alpha$. In particular, $\mathcal{E}$ is the orbital binding energy per unit mass of the generic $\alpha$ particle. The ergodic distribution function corresponding to a spherical density profile can be computed using Eddington's formula \citep[][see also our Appendix \ref{app:diff_coe}]{1916MNRAS..76..572E, 2013degn.book.....M}, which simplifies to the expression
        \begin{equation}\label{eq:f_E}
            f_\alpha(\mathcal{E}) = \frac{1}{\sqrt{8}\pi^2} \int^\mathcal{E}_0 \frac{\mathrm{d}\psi}{\sqrt{\mathcal{E}-\psi}}\frac{\mathrm{d}^2 n_\alpha}{\mathrm{d}\psi^2}
        \end{equation}
        given our assumptions. For a Dehnen density profile, $n_\alpha \propto \psi^4$ at infinite distance, as $n_\alpha \propto r^{-4}$ and $\psi \propto r^{-1}$. Thus,
        \begin{equation}
            \left.\frac{\mathrm{d}n_\alpha}{\mathrm{d}\psi}\right\rvert_{\psi=0} = \left.4 \psi^3\right\rvert_{\psi=0} = 0 \, .
        \end{equation}

        In general, the spherically symmetric distribution function $f$ can depend also on the quantity
        \begin{equation}
                \mathcal{R} (\mathcal{E}, J ) = \frac{J^2}{J^2_\mathrm{c} (\mathcal{E})} \, ,
        \end{equation}
        where $J$ is the magnitude of the specific angular momentum of the orbiting body, and $J_\mathrm{c}$ is its value for a circular orbit of energy $\mathcal{E}$.
        \citet{1978ApJ...226.1087C} showed that
        \begin{equation}
            f(\mathcal{E}, \mathcal{R}) = f_\mathrm{N} (\mathcal{E} ) \frac{\ln \mathcal{R} - \ln \mathcal{R}_\mathrm{q}}{\mathcal{R}_\mathrm{q} -1 - \ln \mathcal{R}_\mathrm{q}}
        \end{equation}
        is a steady-state expression for $f$ near the loss cone. Here $f_\mathrm{N}$ is a normalisation factor, while $\mathcal{R}_\mathrm{q}$ is defined as
        \begin{equation}
            \mathcal{R}_\mathrm{q} (\mathcal{E}) = \mathcal{R}_\mathrm{lc} (\mathcal{E} ) \exp \left(-\sqrt[4]{q^2+q^4} \right) \, .
        \end{equation}
        In this definition, $\mathcal{R}_\mathrm{lc}$ denotes the value of $\mathcal{R}$ at the edge of the loss cone, while $q$ is the loss cone diffusivity, defined as the ratio between the expected relative change in $\mathcal{R}^2$ due to two-body relaxation at the edge of the loss cone:
        \begin{equation}
            q ( \mathcal{E} ) = \left.\frac{\langle \Delta \mathcal{R}^2 \rangle}{\mathcal{R}^2}\right\rvert_{\mathcal{R} = \mathcal{R}_\mathrm{lc}} \, .
        \end{equation}
        The $q$ parameter distinguishes between the full ($q \gg 1$) and the empty ($q \ll 1$) loss cone regimes. In the full loss cone regime, two-body relaxation is particularly efficient in scattering angular momentum, with variations as large as the angular momentum itself. Therefore, objects instantaneously located on an orbit penetrating the loss cone are likely to be scattered away from it before reaching the pericentre, and escape the capture by the MBH. On the other hand, in the empty loss cone regime, once a body enters the loss cone it has very little chance of escaping it, as two-body relaxation induces very small kicks in angular momentum over a single orbit \citep{1978ApJ...226.1087C, 2013degn.book.....M}.

    \subsection{Gravitational wave emission and loss cone captures}\label{sec:EMRI/DP}
        
        The dynamics of a stellar-mass BHs orbiting an MBH in a nuclear cluster is defined by two fundamental timescales:
        \begin{enumerate}
            \item $t_\mathrm{rlx}$, the average time needed for two-body encounters to significantly change the orbital parameters;
            \item $t_\mathrm{GW}$, the average time needed for the emission of GWs to significantly change the orbital parameters.
        \end{enumerate}
        When the stellar-mass BH and the MBH are far enough, the emission of GWs is negligible. In this configuration, the orbit evolution is stochastic and results only from the effect of two-body encounters. These encounters shuffle the position of the orbit on the $(1-e, a)$ plane and are particularly efficient in changing $1-e$. Observing that $\mathcal{R} = 1-e^2$ for Keplerian orbits, we can then define the average time needed for two-body relaxation to significantly change $1-e$ as
        \begin{equation}
            t_\mathrm{rlx} = \frac{1-e}{\mathrm{D_\mathrm{o}}[\Delta (1-e)]} \simeq \frac{1-e^2}{\mathrm{D_\mathrm{o}}[\Delta (1-e^2)]} = \frac{\mathcal{R}}{\mathrm{D_\mathrm{o}}[\Delta \mathcal{R}]} \, ,
        \end{equation}
        where we use the fact that $1-e^2$ goes like $2(1-e)$ for $e \to 1$. The expression we used for $\mathrm{D_\mathrm{o}}[\Delta \mathcal{R}]$ can be found in Appendix \ref{app:diff_coe} and in \citet{2022MNRAS.514.3270B}. 
        %\as{Can't we also write an explicit analytical expression for $t_\mathrm{rlx}$ as a function of the relevant system parameters, the same way we do for $t_{\rm GW}$ in eq 20? I don't see that anywhere in the paper.} \lb{We should add the formula of the coefficient in the appendix, but it can't be written in a simple analytical way as it is contains the integrals over the distribution function. If we want to use explicit analytic expressions we need to fit them like in Bortolas Mapelli 2019 or use one of their estimates like A13 (that are formulas for distribution functions corresponding to single power law profiles and one stellar mass object).}

        During this random motion on the $(1-e, a)$ plane, the orbit might end up in a situation where the stellar-mass BH passes so close to the MBH that the emission of GWs begins to contribute to the evolution of the orbit, pushing it to become tighter and less eccentric, as the system loses energy and angular momentum. We can measure the strength of this effect by looking at the $t_\mathrm{GW}$ timescale, which we define as
        \begin{equation}
            t_\mathrm{GW} = (1-e) \left\langle \frac{\mathrm{d}}{\mathrm{d}t } \left( 1-e \right) \right\rangle^{-1}_\mathrm{GW} = - \left( 1-e \right) \left\langle \frac{\mathrm{d}e}{\mathrm{d}t } \right\rangle^{-1}_\mathrm{GW} \, ,
        \end{equation}
        where angle brackets indicate orbit-averages. \citet{1964PhRv..136.1224P} showed the following:
        \begin{equation}
            \left\langle \frac{\mathrm{d}e}{\mathrm{d}t } \right\rangle_\mathrm{GW} = - \frac{304}{15} \frac{G^3 M_\bullet m_\bullet (M_\bullet+m_\bullet)}{c^5 a^4} \frac{e}{(1-e^2)^{5/2}}\left( 1+\frac{121}{304} e^2 \right) \, .
        \end{equation}
        Thus,\footnote{We point out that the same expression of $t_\mathrm{GW}$ for $e \to 1$ can be found by defining $t_\mathrm{GW} = | \,a / \left\langle \mathrm{d}a/\mathrm{d}t  \right\rangle_\mathrm{GW}$\,|.}
        \begin{equation} \label{eq:t_gw}
            \begin{aligned}
                t_\mathrm{GW} &= \frac{15}{304} \frac{c^5 a^4}{G^3 M_\bullet m_\bullet (M_\bullet+m_\bullet)} \frac{(1-e)(1-e^2)^{5/2}}{e} \left( 1+\frac{121}{304} e^2 \right)^{-1} \\
                & \xrightarrow{e \to 1} \frac{12 \sqrt{2}}{85} \frac{c^5 a^4}{G^3 M_\bullet m_\bullet (M_\bullet+m_\bullet)} (1-e)^{7/2} \, .
            \end{aligned}
        \end{equation}

        At this point, we can define two regions in the $(1-e,a)$ plane (see Fig.\ \ref{fig:plane}):
        \begin{figure}
            \centering
            \resizebox{\hsize}{!}{\includegraphics{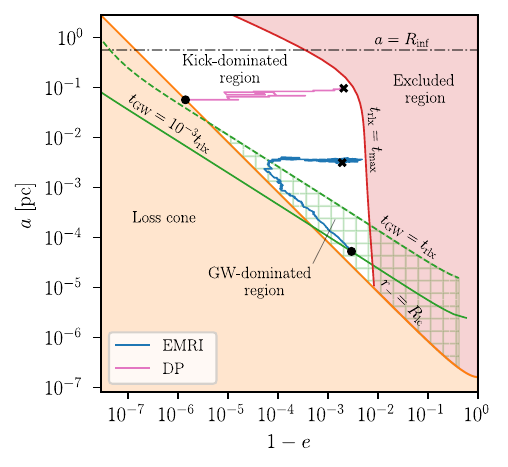}}
            \caption{Two examples of orbit evolution in the $(1-e,a)$ plane. The blue line shows the formation of an EMRI, while the pink one results in a DP.  Both are mainly randomised in eccentricity while inside the white kick-dominated region. When the blue line crosses into the GW-dominated region, which is filled with a green grid, it predominantly evolves via GW emission. DPs are identified by the crossing of the loss cone edge, which is represented by the solid orange curve. EMRIs instead must reach the portion of the solid green curve in between the loss cone and the excluded region (see Sect.\ \ref{sec:stopping}). Here we set $M_\bullet = 3 \times 10^5 \, \mathrm{M}_\sun$.}
            \label{fig:plane}
        \end{figure} 
        \begin{enumerate}
            \item the kick-dominated region, where $t_\mathrm{rlx} < t_\mathrm{GW}$ and the orbit evolution is stochastic and dominated by two-body encounters;
            \item the GW-dominated region, where $t_\mathrm{GW} < t_\mathrm{rlx}$ and the orbit evolution is quasi-deterministic and dominated by the emission of GWs.
        \end{enumerate}
        
        Since strong velocity kicks might still push the orbit back into the kick-dominated region, or directly inside the loss cone after few orbital periods,\footnote{The latter of these could be called EMRIs, as they spend some time in the GW-dominated region before merging. However, given the small number of cycles completed before the merger, the GW signal from these events will be too weak to be observed by LISA. We therefore classify them as DPs.} we caution against identifying the EMRI formation region with the GW-dominated region. Following \cite{2005ApJ...629..362H}, a safest choice is that of labelling as EMRIs the orbits that reach the condition $t_\mathrm{GW} = 10^{-3} t_\mathrm{rlx}$. Here the orbital elements are so deep into the GW-dominated region that we can be confident that a merger will eventually take place after several thousands of periods. We instead define DPs as those events where the orbit enters the loss cone before reaching the $t_\mathrm{GW} = 10^{-3} t_\mathrm{rlx}$ condition.

\subsection{Loss cone definition in post-Newtonian gravity}

        From GR, we know that a test particle on a stable orbit of eccentricity $e$ must always keep a distance from an MBH of at least $(6+2e)/(1+e)$ gravitational radii in order to avoid plunging into it \citep{1994PhRvD..50.3816C}. Consistently, the innermost stable circular ($e=0$) orbit around an MBH has radius $6 R_\mathrm{g}$, while a body on a parabolic ($e=1$) trajectory can approach the MBH up to $4 R_\mathrm{g}$, still managing to escape it \citep{1973grav.book.....M}. Therefore, $4 R_\mathrm{g}$ can be taken as the loss cone radius for very eccentric orbits in GR.

        Relaxation in dense systems, however, is usually modelled with Newtonian gravity, which poses a problem for the loss cone definition. Crucially, there is no single method to translate Keplerian orbits of Newtonian gravity into geodesics of the Schwarzschild metric in GR \citep{2017PhRvD..95h3001S}. In particular, two different mappings can both recover the same Newtonian ellipse for large pericentres. One mapping is that where both the general relativistic and Newtonian orbits share the same pericentre. This mapping has the property of conserving the circumference of circular orbits between the two theories \citep{2017PhRvD..95h3001S}, which is a property that does not appear to be much of use for our study, as we are interested in highly eccentric orbits. Instead, a more useful mapping is the one that preserves the angular momentum between the theories. In this case, we can identify plunging orbits around Schwarzschild MBHs with the same condition used in GR \citep{2011PhRvD..84d4024M,2011CQGra..28v5029S,2012CQGra..29u7001W,2013degn.book.....M}:
        \begin{equation}
            J < \frac{4GM_\bullet}{c} \, .
        \end{equation}
        From the Newtonian relation $J^2 = GM_\bullet a (1-e^2)$, then follows the equivalent condition for the pericentre when $e \to 1$:
        \begin{equation}
            r_- < 8 R_\mathrm{g} \, .
        \end{equation}
        To get our point across let us consider an example. Take a body moving in Newtonian gravity on a highly eccentric orbit, passing at slightly less than $8 R_\mathrm{g}$ from the MBH at pericentre and evolve it back in time to the apocentre of its orbit. Given the high eccentricity we assume, the apocentre is large enough that GR simplifies to Newtonian gravity. If we now turn on GR and let the system evolve forward in time, the same body will reach a minimum distance from the MBH slightly lower than $4 R_\mathrm{g}$, plunging into its horizon. Thus, when simulating similar orbits in Newtonian dynamics, we have to consider as plunges all bodies that get to less than $8 R_\mathrm{g}$ from the MBH.

        However, for this work we neither solved the Einstein field equations to evolve our system nor relied on a Newtonian description of the dynamics. We evolved our systems using PN dynamics, which operates in an intermediate regime by definition. We are not aware of studies on the capture of eccentric orbit at the PN level, unlike in the case of circular orbits \citep[e.g.][]{2024LRR....27....2S}. Thus, we first test how the PN dynamics affects the position of the pericentre for a highly eccentric and deeply penetrating orbit within our code.

        We took a test orbit initialised such that the stellar-mass BH reaches a distance from the MBH of $8 R_\mathrm{g}$ at pericentre in Newtonian dynamics. Then, we ran the same initial conditions for three more times, adding a new PN term each time. We repeated this test for different combinations of masses and eccentricities, and found  that only the latter have some small influence on the result. In particular, we tested for all combinations of $e = \{ \,0.9,0.999,0.99999\, \} $, $m_\bullet=\{\,10,100\,\} \, \mathrm{M_\sun}$, and $M_\bullet/m_\bullet = \{\,10^3,10^4,10^5,4 \times 10^5\,\}$.

\begin{figure}
            \centering
            \resizebox{\hsize}{!}{\includegraphics{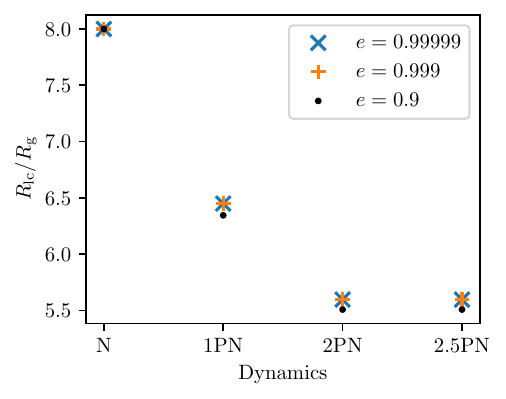}}
            \caption{Loss cone radius per unit of $R_\mathrm{g}$ as a function of $e$ and dynamics used to evolve the system. We also tested for different $M_\bullet$ and $m_\bullet$ values, finding no discernible difference. On the x-axis, `N' stands for Newtonian.}
            \label{fig:losscone}
        \end{figure}

        We find that, regardless of the choice of parameters, a highly eccentric orbit with $r_-=8 R_\mathrm{g}$ in Newtonian dynamics reaches a pericentre of ${\sim}6.45 R_\mathrm{g}$ at 1PN order and a pericentre of ${\sim}5.6 R_\mathrm{g}$ at 2PN and 2.5PN order. The results of this investigation are shown in Fig.\ \ref{fig:losscone}.
        Thus, we define the loss cone radius as
        \begin{equation} \label{eq:R_lc}
                R_\mathrm{lc} = \zeta R_\mathrm{g} \, ,
        \end{equation}
        with
        \begin{equation}
            \zeta = 
                \begin{cases}
                    8 & \text{if Newtonian dynamics} \\
                    6.45 & \text{if 1PN dynamics} \\
                    5.6 & \text{if 2PN or 2.5PN dynamics}
                \end{cases} \, .
        \end{equation}

    \subsection{EMRI-to-DP ratio and cliffhanger EMRIs} \label{sec:cliff_region}

    Through Monte Carlo simulations, \citet{2005ApJ...629..362H} were the first to explore how the probability of EMRI versus DP depends on the initial semi-major axis of the orbit $a_0$.

    The code they presented works in $(\mathcal{E},J)$ space and follows an object on a relativistic orbit, with small perturbations due to GW emission and stochastic two-body scattering. It accounts for GW radiation using \citep{1964PhRv..136.1224P}
        \begin{equation} \label{eq:dE/dt}
            \left\langle \frac{\mathrm{d}\mathcal{E}}{\mathrm{d}t } \right\rangle_\mathrm{GW} = \frac{32}{5} \frac{G^4 M_\bullet^2 m_\bullet (M_\bullet + m_\bullet)}{c^5 a^5 (1-e^2)^{7/2}} \left( 1 + \frac{73}{24} e^2 + \frac{37}{96} e^4 \right)
        \end{equation}
        and
        \begin{equation}
            \left\langle \frac{\mathrm{d}J}{\mathrm{d}t } \right\rangle_\mathrm{GW} = - \frac{32}{5} \frac{G^{7/2}M_\bullet m_\bullet(M_\bullet+m_\bullet)^{3/2}}{c^5a^{7/2}(1-e^2)^2} \left( 1+ \frac{7}{8} e^2 \right) \, ,
        \end{equation}
        which are multiplied by $P$ to compute an orbit-averaged estimate of the energy and angular momentum carried away by GWs during a period. Similarly, the code also updates the angular momentum once per orbit to account for two-body relaxation, by employing the orbit-averaged diffusion coefficients $\mathrm{D_o}[\Delta J]$ and $\mathrm{D_o}\left[\left(\Delta J\right)^2\right]$. It instead neglects scattering in $\mathcal{E}$, as two-body relaxation is much more efficient in shuffling $J$ rather than $\mathcal{E}$ \citep{2008gady.book.....B,2013degn.book.....M}. 
        
        After running many simulations for a central MBH of mass $M_\bullet = 3 \times 10^6 \, \mathrm{M_\sun}$ surrounded by stellar-mass BHs of mass $m_\bullet = 10 \, \mathrm{M_\sun}$, \citet{2005ApJ...629..362H} focused on the function
        \begin{equation}
           S(a_0) = \frac{N_\mathrm{EMRI}(a_0)}{N_\mathrm{EMRI}(a_0)+N_\mathrm{DP}(a_0)} \, ,
        \end{equation}
        where $N_\mathrm{EMRI}$ and $N_\mathrm{DP}$ are respectively the counts of EMRIs and DPs formed given an initial $a_0$, and discovered the existence of a critical value $a_\mathrm{c}^\mathrm{HA} \simeq 5 \times 10^{-3} R_\mathrm{inf}$ such that $S(a_0) = 1$ for $a_0 \ll a_\mathrm{c}^\mathrm{HA}$, and $S(a_0) = 0$ for $a_0 \gg a_\mathrm{c}^\mathrm{HA}$. \citet{2021MNRAS.501.5012R} fixed a sign error in the original implementation of \citet{2005ApJ...629..362H}, updating the result to $a_\mathrm{c}^\mathrm{RP} \simeq 2 \times 10^{-2} R_\mathrm{inf}$.
        
        Recently, QS24 studied again the problem, considering a more realistic set up in which the stellar-mass BH is scattered by stars of mass $m_\star = 1\, \mathrm{M_\sun}$ and scattering in energy is not neglected. Most notably, they extended the study for low-mass MBHs, finding that it is possible for EMRIs to form even from large $a_0$, thus breaking the classical notion of a somewhat sharp transition between $a_0$ values from which only EMRIs form to those that result in DPs only.

\begin{figure}
            \centering
            \resizebox{\hsize}{!}{\includegraphics{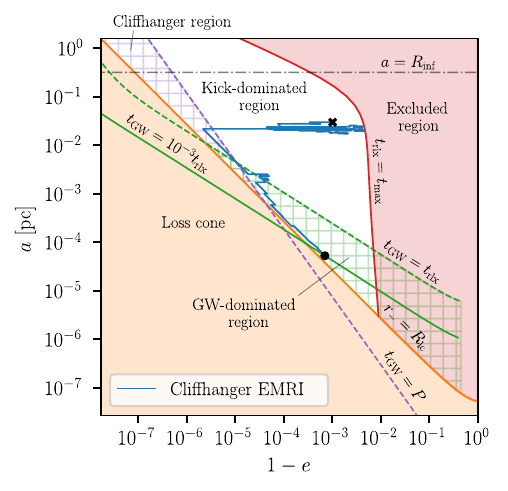}}
            \caption{Example of a cliffhanger EMRI orbit in the $(1-e,a)$ plane. The cliffhanger region is filled with a purple grid, and it is delimited by the purple dashed line and the orange solid curve (see Fig.\ \ref{fig:plane} for the meaning of the other elements). The sharp tightening of the orbit at the passage inside the cliffhanger region occurs during a single pericentre passage. Here we set $M_\bullet = 10^5 \, \mathrm{M}_\sun$.}
            \label{fig:cliff}
        \end{figure}

        \citet{2024PhRvL.133n1401Q} called EMRIs that form from initially wide orbits cliffhanger EMRIs. The name has to do with the suspenseful ending of the orbital evolution, leading to an EMRI in a region of the parameter space where only DPs were originally expected. In the classical picture, stellar-mass BHs on wide orbits generally end up plunging into the loss cone following a sequence of two-body encounters, which culminate in an almost radial orbit towards the MBH. However, a fraction of these orbits will not enter the capture sphere of the MBH, but graze its surface at a slightly larger distance. In this case, the burst of GW radiation emitted is so strong to easily both reduce $a$ and increase $1-e$ by a factor of ten during a single passage (see Fig.\ \ref{fig:cliff}). If the mass of the MBH is small and the semi-major axis large enough, following the large loss of energy and angular momentum, the stellar-mass BH will end up on a much tighter orbit, closer to or inside the GW-dominated region, with a high probability of later becoming an EMRI. 

        To better quantify how small the MBH mass should be and to better characterise the phenomenon, we here introduce the `cliffhanger region': a region on the $(1-e,a)$ plane from which cliffhanger EMRIs come from. Given Eqs.\ \eqref{eq:period} and \eqref{eq:t_gw}, the semi-major axis $a_\mathrm{cliff}$ where $t_\mathrm{GW} = P$ as a function of the eccentricity is
        \begin{equation} \label{eq:a_cliff}
            a_\mathrm{cliff} = \left( \frac{85 \pi}{6 \sqrt{2}} \right)^{2/5} \left( \frac{m_\bullet^2 (M_\bullet + m_\bullet)}{M_\bullet^3} \right)^{1/5} (1-e)^{-7/5} R_\mathrm{g} \, ,
        \end{equation}
        where we assume $e \to 1$. The loss cone boundary, from Eqs.\ \eqref{eq:peri_apo} and \eqref{eq:R_lc}, is instead identified by the equation $r_-=R_\mathrm{lc}$, that is
        \begin{equation} \label{eq:losscone_edge}
            a_\mathrm{lc} = \frac{\zeta}{1-e} R_\mathrm{g} \, .
        \end{equation}
        The region in the $(1-e,a)$ plane identified by the condition
        \begin{equation}
            a_\mathrm{lc} (e) < a < a_\mathrm{cliff} (e)
        \end{equation}
        is what we  refer to as the cliffhanger region (see Fig.\ \ref{fig:cliff}). As the curves cross at eccentricity
        \begin{equation}
            \tilde{e} = 1 - \frac{85 \pi}{6 \sqrt{2}} \zeta^{-5/2} \left( \frac{m_\bullet^2 (M_\bullet + m_\bullet)}{M_\bullet^3} \right)^{1/2} \, ,
        \end{equation}
        the cliffhanger region is defined only for large eccentricities ($e>\tilde{e})$.
        
        Whenever a stellar-mass BH is scattered by two-body encounters onto an orbit which lays inside the cliffhanger region, the orbit will be drastically modified by GW radiation during a single pericentre passage, meaning the variation happens in less than an orbital period.
        The stellar-mass BH will not plunge onto the MBH after this strong emission. During this process the orbital parameters are related through \citep{1964PhRv..136.1224P}
        \begin{equation} \label{eq:a(e)}
            a(e) = c_0 \frac{e^{12/19}}{1-e^2} \left( 1 + \frac{121}{304} e^2 \right)^{870/2299} \, ,
        \end{equation}
        where $c_0$ is determined by the initial conditions. For $e \to 1$, this implies that $a(e) \propto (1-e)^{-1}$. Therefore, GWs emission conserves the pericentre\footnote{This also implies that $J$ is approximately constant, as $J \simeq \sqrt{2GM_\bullet r_-}$ for highly eccentric orbits.} for $e \to 1$, and in the $(1-e,a)$ plane the orbit will evolve parallel to the loss cone edge.\footnote{Hence the EMRI sort of `hangs on the cliff' at the edge of the loss cone, re-proposing the pun at the origin of the title of the movie `Cliffhanger' (1993).}

\begin{figure}
            \centering
            \resizebox{\hsize}{!}{\includegraphics{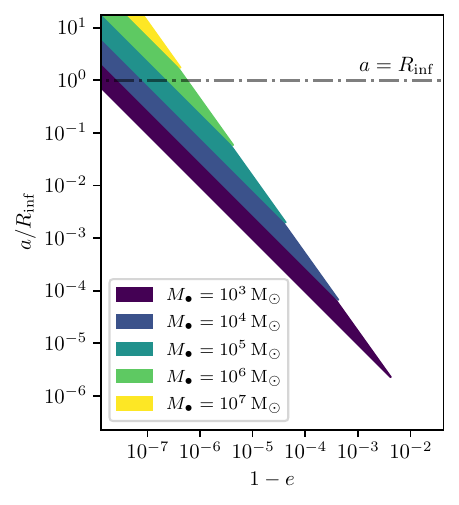}}
            \caption{Cliffhanger region for different MBH masses in the $(1-e, a)$ plane, where $a$ is shown through its ratio with the influence radius. The darker regions are partly covered by lighter regions as all the regions extend up to the upper axis. All regions share a similar triangular shape, which is pushed towards the upper left corner of the plot as $M_\bullet$ increases, effectively shrinking the available cliffhanger region for $a < R_\mathrm{inf}$. For $M_\bullet = 10^7 \, \mathrm{M}_\sun$ the whole cliffhanger region is located outside the $a=R_\mathrm{inf}$ line.}
            \label{fig:cliff_masses}
        \end{figure}

        The scaling of the two curves with the mass of the central MBH is different: Eq.\ \eqref{eq:losscone_edge} implies that $a_\mathrm{lc} \propto M_\bullet$, while Eq.\ \eqref{eq:a_cliff} shows that $a_\mathrm{cliff} \propto M_\bullet^{3/5}$. As we consider larger MBHs, the loss cone edge will cross the $t_\mathrm{GW} = P$ curve at larger eccentricities and semi-major axes, eventually to $a>R_\mathrm{inf}$ (see Fig.\ \ref{fig:cliff_masses}). As a result, it gets increasingly harder with larger values of $M_\bullet$ for two-body relaxation to scatter orbits inside the cliffhanger region while avoiding DPs, as the corresponding range in $e$ becomes extremely small.
        We can therefore predict the maximum MBH mass which allows   the formation of cliffhanger EMRIs as in QS24. From Eqs.\ \eqref{eq:period} and \eqref{eq:dE/dt}, follows that the change in energy in a period is
        \begin{equation} \label{eq:DeltaE_GW}
            \begin{aligned}
                \Delta \mathcal{E}_\mathrm{GW} (a,e) &= P \left\langle \frac{\mathrm{d}\mathcal{E}}{\mathrm{d}t } \right\rangle_\mathrm{GW} \\ &\xrightarrow{e \to 1} \frac{85 \pi}{12 \sqrt{2}} \frac{G^{7/2} M_\bullet^2 m_\bullet (M_\bullet + m_\bullet)^{1/2}}{c^5} a^{-7/2} (1-e)^{-7/2} \, ,              
            \end{aligned}
        \end{equation}
        which depends only on the pericentre in the $e\to1$ limit.
        Plugging Eq.\ \eqref{eq:a_cliff} into Eq.\ \eqref{eq:DeltaE_GW} gives
        \begin{equation}
            \Delta \mathcal{E}_\mathrm{GW}^\mathrm{cliff} \big(a_\mathrm{cliff},e(a_\mathrm{cliff})\big) = \frac{GM_\bullet}{2 a_\mathrm{cliff}}
        \end{equation}
        which is the energy change in a period due to GWs emission on the $t_\mathrm{GW} = P$ curve. On this curve, GW radiation doubles $\mathcal{E}$ during a single period,
        \begin{equation} \label{eq:double}
            \mathcal{E}_\mathrm{f}^\mathrm{cliff} = \mathcal{E}_\mathrm{i}^\mathrm{cliff} + \Delta \mathcal{E}_\mathrm{GW}^\mathrm{cliff} \left(a_\mathrm{cliff,i},e(a_\mathrm{cliff,i})\right)  = \frac{GM_\bullet}{a_\mathrm{cliff,i}} = 2 \mathcal{E}_\mathrm{i}^\mathrm{cliff} \, ,
        \end{equation}
        meaning that $a$ is halved in the process.

        Going deeper into the cliffhanger region allows   larger energy jumps. At the limit of the loss cone edge, $r_- = R_\mathrm{lc}$ and Eq.\ \eqref{eq:DeltaE_GW} gives
        \begin{equation}
            \Delta \mathcal{E}_\mathrm{GW}^\mathrm{lc} \big(a_\mathrm{lc},e(a_\mathrm{lc})\big) = \frac{85 \pi}{12 \sqrt{2}} \zeta^{-7/2} c^2 m_\bullet (M_\bullet + m_\bullet)^{1/2} M_\bullet^{-3/2} \, ,
        \end{equation}
        which does not depend on $a_\mathrm{lc}$ as we are setting the value of the pericentre. On the edge of the loss cone, the energy jump will be large, thus we can approximate
        \begin{equation} \label{eq:energy_approx}
            \mathcal{E}_\mathrm{f}^\mathrm{lc} = \mathcal{E}_\mathrm{i}^\mathrm{lc} + \Delta \mathcal{E}_\mathrm{GW}^\mathrm{lc} \simeq \Delta \mathcal{E}_\mathrm{GW}^\mathrm{lc} \, ,
        \end{equation}
        from which follows that all the cliffhanger jumps on the loss cone edge will end up with the same semi-major axis:
        \begin{equation} \label{eq:a_lc_final}
            a_\mathrm{f}^\mathrm{lc} = \frac{6\sqrt{2}}{85 \pi} \zeta^{7/2} \frac{G}{c^2} \frac{M_\bullet^{5/2}}{m_\bullet(M_\bullet+m_\bullet)^{1/2}} \simeq \frac{6\sqrt{2}}{85 \pi} \zeta^{7/2} \frac{M_\bullet}{m_\bullet} R_\mathrm{g} \, .
        \end{equation}

        Finally, we can get a conservative estimate for the maximum MBH mass expected to produce cliffhanger EMRIs. The burst of GWs will end up forming an EMRI as long as the jump in $a$ is inside or close to the EMRI formation region. Using the classical estimate consistent with our results (see Sect.\ \ref{sec:main_res}), $a_\mathrm{f}^\mathrm{lc} \lesssim 10^{-2} R_\mathrm{inf}$, we obtain the following:
        \begin{equation}
            M_\bullet \lesssim \frac{17 \pi}{120 \sqrt{2}} \zeta^{-7/2} \frac{c^2}{\sigma^2}  m_\bullet \,.
        \end{equation}
        Given our choice of parameters for the M--$\sigma$ relation \citep{2009ApJ...698..198G}, we finally get
        \begin{equation} \label{eq:M_max}
            M_\bullet \lesssim 1.8 \times 10^7 \zeta^{-2.38} \left( \frac{m_\bullet}{10 \, \mathrm{M}_\sun} \right)^{0.68} \, \mathrm{M}_\sun \,.
        \end{equation}
        It should be noted that this derivation takes into account only the proprieties of the MBH and the stellar-mass BHs, while the distributions of background bodies only marginally enters in the effective critical radius $10^{-2} \, R_\mathrm{inf}$. In particular, it does not factor in how likely is for two-body relaxation to push the orbit inside the cliffhanger region, nor the fact that a cliffhanger EMRIs can be the result of more than one energy jump, as it is possible for the orbit to enter the cliffhanger region multiple times.
        
        Plugging $\zeta=5.6$ and $m_\bullet = 10 \, \mathrm{M}_\sun$ into Eq.\ \eqref{eq:M_max}, we get $M_\bullet \lesssim 3 \times 10^5\, \mathrm{M}_\sun$, which is indeed roughly the MBH mass for which we start seeing cliffhanger EMRIs forming (see Sect.\ \ref{sec:main_res}).
        For $\zeta=8$ and $m_\bullet = 10\, \mathrm{M}_\sun$, we find $M_\bullet \lesssim 1.3 \times 10^5 \,\mathrm{M}_\sun$, which is remarkably close to the $10^5\, \mathrm{M}_\sun$ value obtained by QS24, despite our different assumptions. 
        
%__________________________________________________________________

\section{Numerical simulations of EMRI and DP formation} \label{sec:methods}

    In this section we illustrate the features of our numerical implementation. We begin with an overview of the integration of the Hamiltonian evolution, then we explain how our treatment of local and orbit-averaged two-body relaxation works. Finally, we detail our procedure to generate the initial conditions for the runs we perform, together with the stopping criteria we use to identify the formation of EMRIs and DPs.

    \subsection{Numerical overview and setup} \label{sec:setup}
        \begin{figure*}
            \centering
            \resizebox{\hsize}{!}{\includegraphics{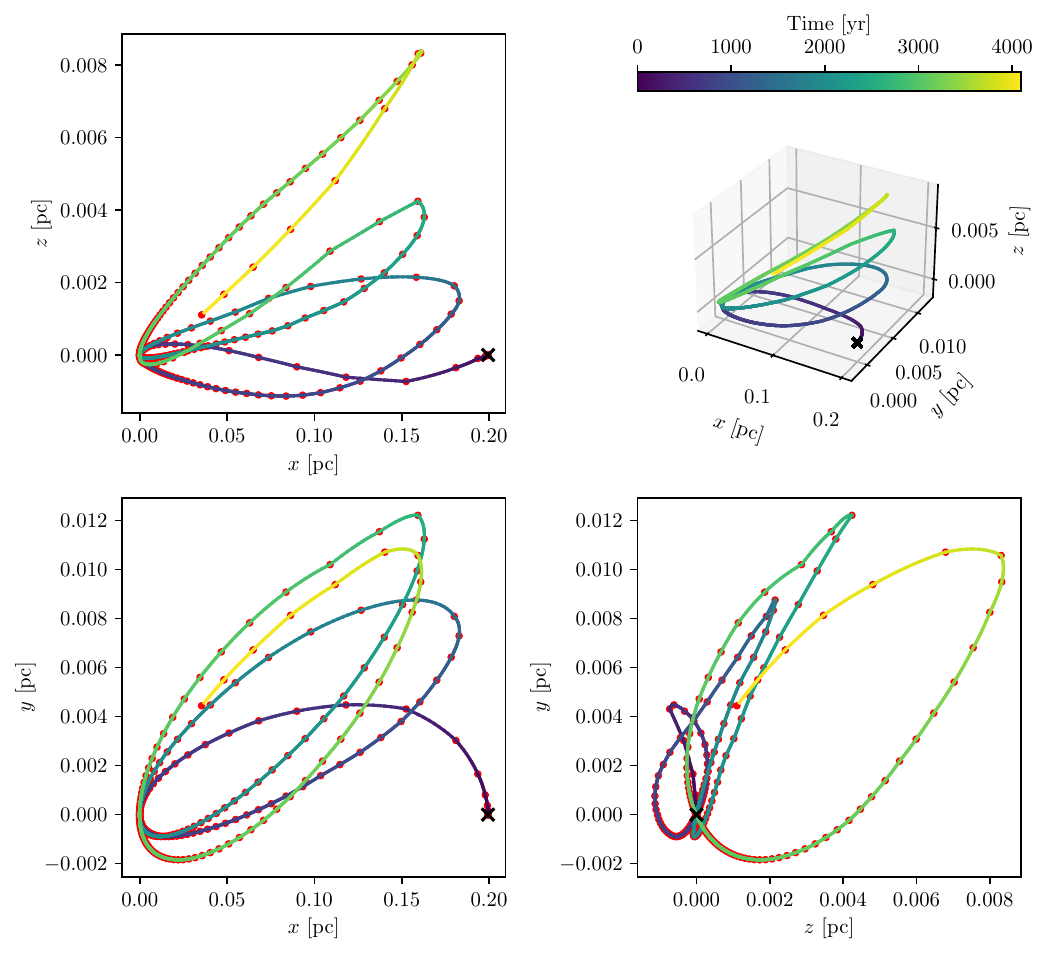}}
            \caption{Evolution of the position of a stellar-mass BH over few passages of the apocentre, using our local treatment of two-body relaxation. The top right panel shows the 3D trajectory of the stellar-mass BH, while the other three panels show its projections onto the three planes of the reference system. The colour of the curve shows the time elapsed since the beginning of the evolution. The red dots indicate the points  where a velocity kick is given and they coincide with the time step of the main integrator. The points  between the red dots are interpolated. The black cross shows the initial position of the stellar-mass BH, while the MBH is fixed at $(0,0,0)$. We note that the $y$- and $z$-axes have different scales for better readability. In this example $M_\bullet = 4 \times 10^6 \, \mathrm{M}_\sun$, $a_0 = 0.1 \, \mathrm{pc}$, $e_0=0.999$.}
            \label{fig:orbits_local}
        \end{figure*}

        In this work we employ an extended version of the code presented in \citet{2016MNRAS.461.4419B}, a PN few-body integrator with adaptive step size, which we use to study the evolution of a binary system comprised of an MBH and a stellar-mass BH. The integrator leverages on a C\texttt{++} implementation of the Bulirsch-Stoer algorithm \citep{Bulirsch1966} coupled with Richardson extrapolation \citep{1911RSPTA.210..307R} to increase the accuracy of the numerical solution. The adaptive time step allows us to get smaller steps when the dynamics is faster, such as at the pericentre of the orbit.

        The code integrates the equation of motion of two-body systems in the Hamiltonian form:
        \begin{equation} \label{eq:hamiltonian}
            H = H_0 + \frac{1}{c^2} H_1 + \frac{1}{c^4} H_2 + \frac{1}{c^5} H_{2.5} + \mathcal{O} \left( \frac{1}{c^6} \right) \, .
        \end{equation}
        Here, $H_0$ is the non-relativistic Newtonian Hamiltonian, while $H_1$, $H_2$ and $H_{2.5}$ are corrective terms introducing purely relativistic effects. Complete expressions for these terms, matching the form implemented numerically, can be found for example in \citet{2016MNRAS.461.4419B}.\footnote{\citet{2016MNRAS.461.4419B} focus on the three-body problem, and therefore report three-body PN Hamiltonian terms. We consistently use the same Hamiltonian, and reduce the evolution to a two-body problem by placing a third body with negligible mass at very large distance from the inner binary, effectively isolating its dynamics.}
        $H_{2.5}$ is the first non-conservative term and accounts for GW radiation \citep{1997PhRvD..55.4712J,2014LRR....17....2B}. We only consider Schwarzschild BHs, thus we ignore the 0.5PN and 1.5PN terms of the expansion, as they account for spin and spin-orbit effects \citep{2024LRR....27....2S}.
        
        We use the Hamiltonian $H$ to evolve the position $\vec{x}$ and the momentum $\vec{p}$ of the stellar-mass BH by integrating its Hamilton's equations in time $t$:
        \begin{equation}
            \begin{cases}
                    \displaystyle \frac{\mathrm{d} \vec{x}}{\mathrm{d} t} = \sum_n \frac{1}{c^{2n}} \frac{\partial H_n}{\partial \vec{p}} \\
                    \displaystyle \frac{\mathrm{d} \vec{p}}{\mathrm{d} t} = - \sum_n \frac{1}{c^{2n}} \frac{\partial H_n}{\partial \vec{x}} 
            \end{cases} \, .
        \end{equation}
        At the same time we forcefully keep the MBH fixed and still at the origin of our Cartesian coordinate system. This choice is motivated by the fact that in a nuclear cluster the MBH will simultaneously interact with virtually all the objects in the stellar distribution, which will only cause a small Brownian motion-like displacement from the centre.
        %and its position will be, on average, fixed at the centre of the stellar distribution. 
        We ignore such a stochastic displacement of the MBH, even if it might be relevant as the number of stellar objects and the mass of the MBH gets lower \citep{2013degn.book.....M}.
        Together with the evolution of the binary system, the code can account for the presence of additional external gravitational potentials, possibly representing a distribution of background bodies. We note again that only a single stellar-mass BH is evolved by directly solving Hamilton's equations.
        
        In this work, we extend the code to include the effects of two-body relaxation on the orbit of the stellar-mass BH. These are due to the gravitational interaction of the stellar-mass BH with the rest of the stellar distribution. These stochastic interactions are fully characterised by the diffusion coefficients of the velocity component computed from their distribution function \citep{shapiroStarClustersContaining1978,1978ApJ...226.1087C,2013degn.book.....M}. We include the effects of two-body relaxation via a Monte Carlo approach by repeating the following procedure at the end of each time step of the integration:
        \begin{enumerate}
            \item we compute the mean and the standard deviation of the expected variation $\langle \Delta \vec{v}\rangle$ of the velocity $\vec{v}$ of the stellar-mass BH, considering the two-body encounters with the scattering objects which happened during the last time step;
            \item we randomly draw a velocity change $\Delta \vec{v}$ from the appropriate Gaussian distribution, which we build according to the previous step;
            %such that the expected change of $\vec{v}$ is recovered over many random extractions from that distribution;
            \item before moving to the next time step, we update the momentum of the stellar-mass BH to $\vec{p'} = \vec{p} + m_\bullet \Delta \vec{v}$, where $m_\bullet$ is the mass of the stellar-mass BH.\footnote{Here we use the relation $\vec{p} = m \vec{v}$, which is not valid at PN order. Thus, while the evolution of the system is PN, the small velocity kicks we give to the stellar-mass BH at the end of each time step are Newtonian.
            %\as{Is this justifiable? Is it possible to estimate whether this matters or not?}\lb{I think the comparison should be to compare $m \delta v$ with $m \, \partial_v (\gamma v) \simeq \gamma m \delta v$. The maximum speed is at the pericentrer $c/\sqrt{\zeta}=c/2$. Quindi sottostimiamo il kick al massimo del $16\%$.} \dm{I add that $\Delta v / v$ is at most $\sim 10^{-2}$ with occasional peaks larger than that for some cases I tested (different masses and outcomes)}
            }
        \end{enumerate}

        Following this procedure, we can statistically account for two-body encounters locally and multiple times within a single orbit, without leveraging on orbit-by-orbit approaches (QS24) or orbit-averages of the diffusion coefficients \citep{zhangMonteCarloStellarDynamics2023a}.

        For our simulations, we consider a stellar-mass BH of mass $m_\bullet = 10 \, \mathrm{M}_\sun$ and explore five possible masses for the MBH: $M_\bullet =  \{ \,10^4, 10^5, 3 \times 10^5, 10^6, 4 \times 10^6\,\} \,\mathrm{M}_\sun$. We also group the background objects into two populations:
        \begin{enumerate}
            \item stars with $m_\star = 1\, \mathrm{M}_\sun$, $M_{\mathrm{tot},\star} = 20 M_\bullet$, $r_{\mathrm{a},\star} = 4 R_\mathrm{inf}$ and $\gamma_\star = 1.5$;
            \item stellar-mass BHs with $m_\bullet = 10\, \mathrm{M}_\sun$, $M_{\mathrm{tot},\bullet} = 0.2 M_\bullet$, $r_{\mathrm{a},\bullet} = 4 R_\mathrm{inf}$ and $\gamma_\bullet = 1.7$.
        \end{enumerate}
        This choice corresponds to the two-components Bahcall-Wolf solution commonly used in literature \citep[][$\gamma \simeq 1.5$ for the lightest component, $\gamma \simeq 1.75$ for the heaviest component]{bahcallStarDistributionMassive1977}, where we slightly reduced the slope of the heaviest component to better represent their distribution in the inner regions of the stellar cluster, where they dominate relaxation \citep[see Fig.\ 4 of][for the quasi-relaxed distribution of the stellar-mass BHs]{2022MNRAS.514.3270B}. The scale radius of the distribution $r_a$ is set so that the stellar cluster in which our $4\times 10^6\mathrm{M}_\sun$ MBH is immersed is consistent with observational estimates of the density at the influence radius of SgrA$^*$ \citep[e.g.][]{2007A&A...469..125S}.
    
    \subsection{Local treatment of two-body relaxation} \label{sec:local}

        As explained in Appendix \ref{app:diff_coe}, in order to compute local diffusion coefficients we have to first find $f_\alpha(\mathcal{E})$ given $n_\alpha$ and then integrate it over $\mathcal{E}$. This process involves nested integrals and derivatives, making it impractical to numerically perform such computations on the fly at each time step of our simulations.
        Thus, we precompute the auxiliary functions $\mathcal{I}_{k,\alpha}$ and $\mathcal{J}_{k,\alpha}$, that are directly related to the diffusion coefficients, for some logarithmically spaced values of $\mathcal{E}$ and $\psi$, and then, during the execution, we bilinearly interpolate these values of $\mathcal{I}_{k,\alpha}$ and $\mathcal{J}_{k,\alpha}$. In this way, we get approximate local values for those coefficients given the exact position and velocity of the stellar-mass BH at a particular time step.

        The diffusion coefficients in velocity are then directly related to the parameters of the Gaussian distributions characterising $\Delta v_\parallel$ and a $\Delta v_\perp$ at each time step of the integration: 
        \begin{equation}
            \Delta v_\parallel = \mu_\parallel + s_\parallel \sigma_\parallel \, , \qquad \Delta v_\perp = \mu_\perp + s_\perp \sigma_\perp \, .        
        \end{equation}
        Here, $s_\parallel$ and $s_\perp$ are two uncorrelated random values, each extracted from a Normal distribution, while
        \begin{equation}
            \begin{alignedat}{2}
                \mu_\parallel &= \mathrm{D}\left[\Delta v_\parallel\right] \Delta t \, , &\qquad \sigma^2_\parallel &= \mathrm{D}\left[\left(\Delta v_\parallel\right)^2\right] \Delta t - \left(\mathrm{D}\left[\Delta v_\parallel\right] \Delta t \right)^2 \, , \\
                \mu_\perp &= 0 \, , &\qquad \sigma^2_\perp &= \mathrm{D}\left[\left(\Delta v_\perp\right)^2\right] \Delta t \, .
            \end{alignedat}
        \end{equation}
        In these equations, $\Delta t$ is the length of the particular time step we are considering. Thus, the values of $\Delta v_\parallel$ and $\Delta v_\perp$ we draw represent a statistical realisation of the change in velocity due to all of the encounters that have occurred during that time step.
    
        Once $\Delta v_\perp$ is drawn, we set its direction randomly, as it is distributed uniformly in the plane perpendicular to the vector $\vec{v}$. Finally, we vectorially sum together $\Delta v_\parallel$ and $\Delta v_\perp$ to get the total variation $\Delta \vec{v}$ of the velocity of the stellar-mass BH, and update its momentum to $\vec{p'} = \vec{p} + m_\bullet \Delta \vec{v}$ before the next step of the integration of Hamilton's equations begins. Figure \ref{fig:orbits_local} shows how this procedure affects a few consecutive apocentre passages. Deviations from regular elliptical orbits are evident, albeit small. Local velocity kicks are more frequent near pericentre due to the adaptive nature of the time step used by the main integrator.
    
    \subsection{Orbit-averaged treatment of two-body relaxation} \label{sec:OA}
        \begin{figure}
            \centering
            \resizebox{\hsize}{!}{\includegraphics{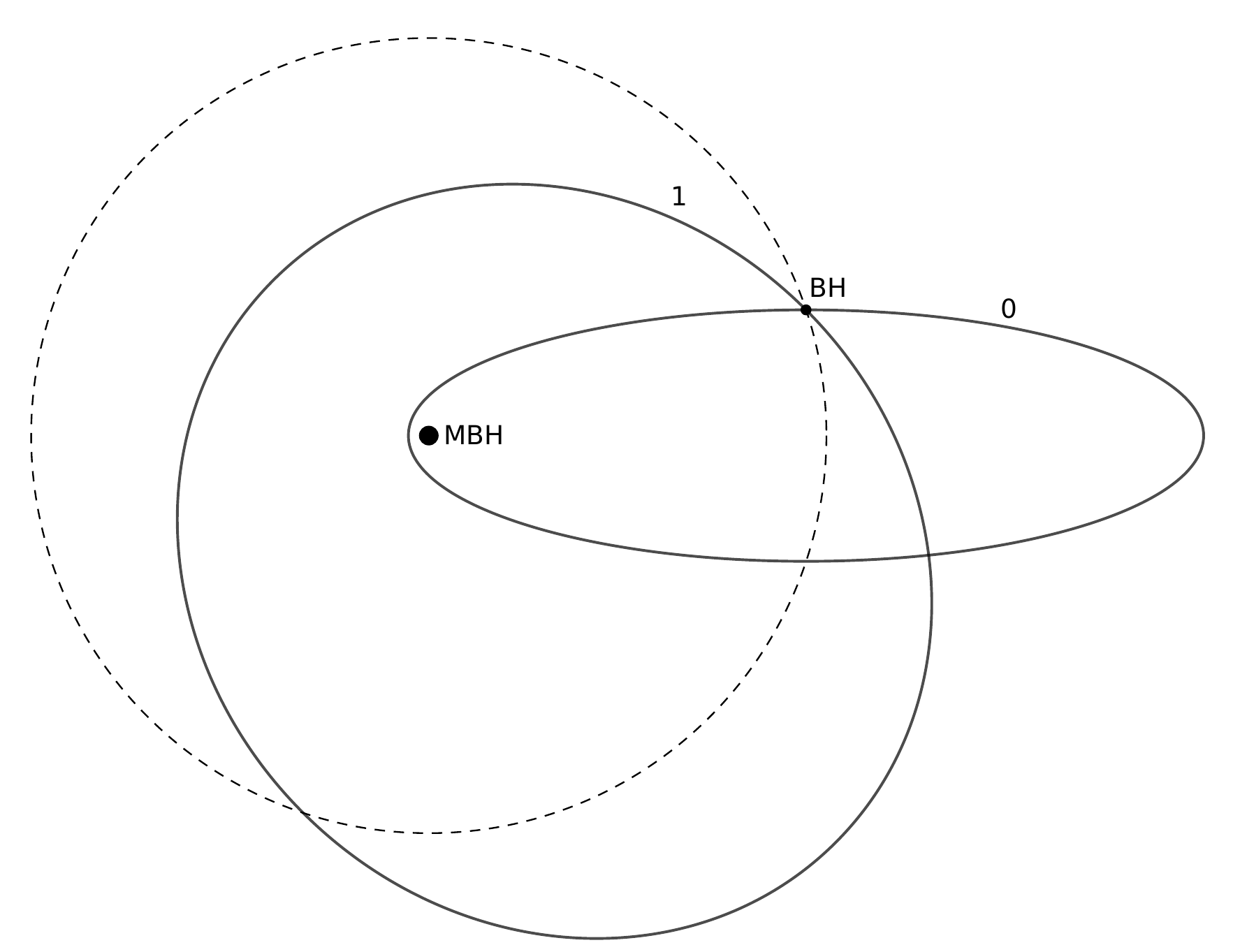}}
            \caption{Sketch depicting an example orbit change following our orbit-averaged procedure. The stellar-mass BH, initially on the orbit labelled  $0$, is kicked when $r=a$, and ends on the orbit labelled  $1$. Both orbits have the same semi-major axis $a$, but different eccentricities. The dashed circle represents a circular orbit with radius $a$. In this example there is no way to rotate either orbit around the focus occupied by the MBH such that the pericentre or apocentre of orbit 0 crosses  orbit 1 at any point. Thus, neither pericentres nor apocentres are good spots to employ our orbit-averaged procedure. We note that the eccentricities of these example orbits are much lower compared to the ones investigated in this work. Moreover, the difference between subsequent orbits is much exaggerated here.}
            \label{fig:sketch}
        \end{figure}
        \begin{figure*}
            \centering
            \resizebox{\hsize}{!}{\includegraphics{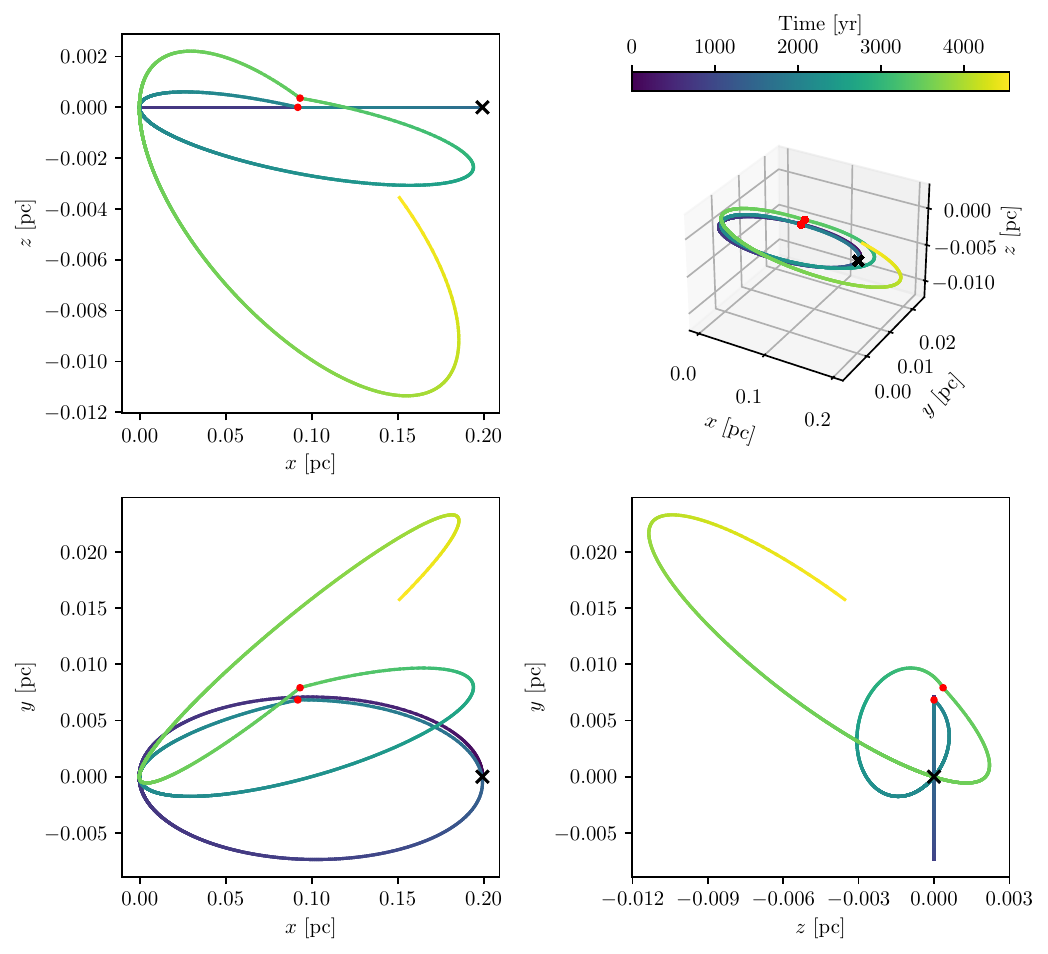}}
            \caption{Evolution of the position of a stellar-mass BH over a few orbits with the orbit-averaged treatment of two-body relaxation. The top right panel shows the 3D trajectory of the stellar-mass BH, while the other three panels show its projections onto the three planes of the reference system. The colour of the curve shows the time elapsed since the beginning of the evolution. The red dots indicate the points where a velocity kick is given, close to $r=a$. We let the body complete the first orbit without applying  kicks to visualise its trajectory better. The black cross shows the initial position of the stellar-mass BH, while the MBH is fixed at $(0,0,0)$. We note that the $y$- and $z$-axes have different scales for better readability. In this example $M_\bullet = 4 \times 10^6 \, \mathrm{M}_\sun$, $a_0 = 0.1 \, \mathrm{pc}$, $e_0=0.997$.}
            \label{fig:orbits_OA}
        \end{figure*}
        The usual approach to implement two-body relaxation into the orbit evolution differs from what we have just described. Codes usually work in the $(\mathcal{E}, J)$ space and thus employ orbit-averaged diffusion coefficients of $\mathcal{E}$ and $J$, which can be computed from those of velocity which we present in Appendix \ref{app:diff_coe} \citep[see][]{2008gady.book.....B,2013degn.book.....M}.
        In this work, we are interested in testing the reliability of the orbit-averaging approximation in opposition to our local approach. To make a meaningful comparison between the two methods, we present some simulations using the orbit-averaged approximation in Sect.\ \ref{sec:tests}.

        Including orbit-averaging in our code is not straightforward, as we do not work with $\mathcal{E}$ and $J$. Here we summarise our implementation:
        \begin{enumerate}
            \item at each time step of the integration, we draw $\delta v_\parallel$ and $\delta v_\perp$\footnote{Here we introduce a new symbol: $\delta$. $\Delta \vec{v}$ is the kick imparted to the stellar-mass BH, while $\delta \vec{v}$ is the randomly drawn change in velocity over the intermediate steps along one orbit.} as explained in Sect.\ \ref{sec:local};
            \item from $\delta v_\parallel$ and $\delta v_\perp$, we compute the corresponding change in energy $\delta \mathcal{E}$ and angular momentum $\delta J$;
            \item we move to the next time step without applying the kick and let the integration proceed, keeping track of the cumulative change in energy $\Delta \mathcal{E} = \sum_i \delta \mathcal{E}_i$ and angular momentum $\Delta J = \sum_i \delta J_i$;
            \item once per orbit, from $\Delta \mathcal{E}$ and $\Delta J$ we compute a compatible kick $\Delta \vec{v}$ and update $\vec{p}$ to $\vec{p'} = \vec{p} + m_\bullet \Delta \vec{v}$;
            \item once the velocity of the stellar-mass BH is changed, we reset $\Delta \mathcal{E}$ and $\Delta J$ to zero, and repeat steps 1-4.     \end{enumerate}        
        Following this procedure, we include the effects of stochastic kicks on the orbit once per period. The changes in $\mathcal{E}$ and $J$ are naturally orbit-integrated as the result of this procedure: the $\delta \mathcal{E}_i$ and $\delta J_i$ terms we sum together are indeed intrinsically weighted by the length of the time steps $\delta t_i$, which contribute through the $\delta v_{\parallel, i}$ and $\delta v_{\perp, i}$ terms.
        
        In Appendix \ref{app:VtoEJ} we show that, once $\delta v_\parallel$ and $\delta v_\perp$ are drawn, then
        \begin{align}
            \delta \mathcal{E} &= -v \delta v_\parallel - \frac{1}{2} \left(\delta v_\parallel\right)^2 - \frac{1}{2} \left(\delta v_\perp\right)^2 \, , \\
            \frac{\delta J}{J} &=
            \begin{aligned}[t]
                 &\frac{\delta v_\parallel}{v} + \delta v_\perp \frac{v_\mathrm{r}}{v v_\mathrm{t}} \sin{\theta} + \frac{1}{2} \left(\frac{\delta v_\perp \cos{\theta}}{v_\mathrm{t}}\right)^2 + \mathcal{O} \left( (\delta v)^3 \right) \, .             
            \end{aligned}
        \end{align}
        Here $v_\mathrm{t} = J/r$ is the tangential velocity, $v_\mathrm{r} = \sqrt{v^2 - v_\mathrm{t}^2}$ is the radial velocity, and $\theta$ is an angle uniformly drawn from the interval $[0, 2 \pi)$. We compute $J$ as
        \begin{equation}
            J = |\vec{J}| = \frac{\vec{x} \times \vec{p}}{m_\bullet} \, .
        \end{equation}
        In Appendix \ref{app:EJtoV} we show that a suitable orbit-integrated kick in velocity can be computed from $\Delta E$ and $\Delta J$ as
        \begin{align} \label{eq:delta_vr}
            \Delta v_\mathrm{r} &= \sqrt{v^2 - 2 \Delta \mathcal{E} - \left(\frac{J + \Delta J}{r}\right)^2} - v_\mathrm{r} \, , \\
            \Delta v_\mathrm{t} &= \frac{(J + \Delta J) \cos{\beta}}{r} - v_\mathrm{t} \, , \\
            \Delta v_\mathrm{3} &= \frac{(J + \Delta J) \sin{\beta}}{r} \, .
        \end{align}
        Here the subscript 3 indicates the direction perpendicular to the orbital plane (thus, $v_\mathrm{3} = 0$ before each kick), while $\beta$ is an angle uniformly drawn from the interval $[0, 2 \pi)$.

        When updating the orbit, we change the velocity of the stellar-mass BH without displacing it. This is possible only if the old and the new orbit intersect at the location where $\Delta v$ is applied, a fact that we ensure by applying the kick when $r\simeq a$ (Fig.\ \ref{fig:sketch} illustrates the situation with a sketch). It is safe to assume that the orbits intersect there, as long as the energy of the particle (and thus $a$, since $\mathcal{E} \simeq G M_\bullet / 2 a$) does not change drastically along a single period.\footnote{We note that the $\Delta \mathcal{E}$ and $\Delta J$ we are using here only come from the effects of two-body relaxation. GWs emission can strongly and quickly modify the energy of the orbit, but their effect is only included in the 2.5PN term of the Hamiltonian.} In case this assumption is not verified, the orbit-averaging approximation is not valid any more, as the time needed for two-body relaxation to significantly change $\mathcal{E}$ is less than the orbital period \citep{2008gady.book.....B,2013degn.book.....M}.
         
        This motivates us to apply the kick $\Delta \vec{v}$ to the stellar-mass BH once per orbit at the time step when $r$ becomes smaller than $a$, coming from the apocentre. This choice reduces the number of cases where Eq.\ \eqref{eq:delta_vr} has no real solutions to a negligible fraction, where orbit-averaging would fail because of the significant change received in one single orbit. In this case, the code simply halts, and we remove the run from the analysis.

        The code computes $a$ and $e$ as
        \begin{equation}
            a = \frac{r_+ + r_-}{2} \, , \qquad e = \frac{r_+ - r_-}{r_+ + r_-} \, ,
        \end{equation}
        each time the pericentre or apocentre of an orbit is reached, where these are identified as the distance from the centre when the radial velocity flips sign. If $a$ and $e$ are computed at apocentre, then $r_-$ is taken as the last pericentre passed. Instead, if they are computed at pericentre, then $r_+$ is taken as the last apocentre passed. 
        Figure \ref{fig:orbits_OA} shows how this procedure affects a few consecutive orbits. In the figure, orbits are almost perfect ellipses, with small deviations due to PN effects.

    \subsection{Stopping criteria and initial conditions} \label{sec:stopping}

        The main aim of the code is that of identifying the formation of EMRIs and DPs. To this scope, we employ the definitions presented in Sect.\ \ref{sec:EMRI/DP}, with some corrections needed given the computational cost of our hybrid few-body/Monte Carlo code.

        As our aim was to gather a statistically significant number of EMRIs and DPs, we chose to limit the parameter space that we explored to a sub-portion close to the loss cone edge. Moreover, as we assume a steady-state stellar profile, we are interested in exploring orbital parameters for which the relaxation time $t_\mathrm{rlx}$ is not too large, as the initial conditions and the diffusion coefficients would be affected by the overall relaxation of the system. In particular, we consider as our timescale of investigation ${\sim}1\%$ of the time-to-peak of (classical) EMRI rates in the systems we are considering, which estimates the timescale of evolution of the distribution function of scatterers \citep{2022MNRAS.514.3270B}
        \begin{equation}
            t_\mathrm{max} = 10 \left( \frac{M_\bullet}{4 \times 10^6\, \mathrm{M}_\sun} \right)^{1.29} \,\mathrm{Myr} \, ,
        \end{equation}
        and exclude from our investigations the region in the $(1-e,a)$ phase space where $t_\mathrm{rlx} > t_\mathrm{max}$. The scaling with $M_\bullet$ for $t_\mathrm{max}$ comes from the fact that \citep{2013degn.book.....M}
        \begin{equation}
            t_\mathrm{rlx} \propto M_\bullet ^2 \sigma^{-2} \propto M_\bullet ^ {1.29} \, ,
        \end{equation}
        where we leverage on the  M--$\sigma$ relation \citep{2009ApJ...698..198G} to obtain the final scaling with the MBH mass. We note that $t_\mathrm{max}$ is not to be intended as a dynamical stopping condition (meaning the simulations do not halt at the time $t=t_\mathrm{max}$), rather the run is ended only when the orbital elements cross over the $t_\mathrm{rlx} = t_\mathrm{max}$ curve on the $(1-e,a)$ plane.

        We initialise all our simulations inside the region of interest where $t_\mathrm{rlx} < t_\mathrm{max}$ following a procedure described at the end of this section and we stop and discard any run where the orbit leaves the region. This choice does not affect the EMRI-to-DP ratio. In fact, $t_\mathrm{max}$ is so long with respect to the orbital period that any plunge or EMRI must enter this region before the capture. As long as the procedure is not biased towards EMRI or DP candidates, once a particle is discarded we are safe to reinitialise it as an uncorrelated particle.

\begin{figure}
            \centering
            \resizebox{\hsize}{!}{\includegraphics{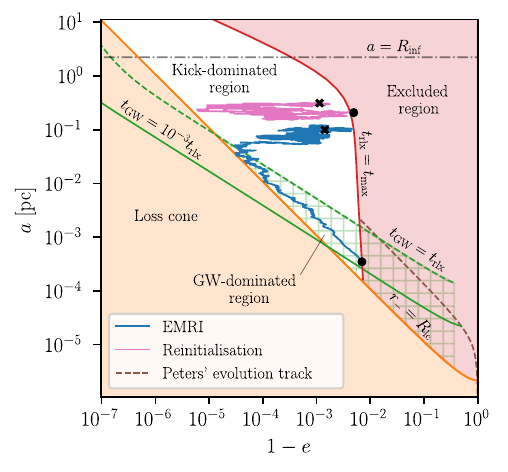}}
            \caption{Two examples of orbit evolution in the $(1-e,a)$ plane. The blue line shows the formation of an EMRI that is missed by our standard definition, and requires the correction described in the text to be rightfully counted. The pink line instead shows a regular reinitialisation. We also plot  the evolution line of an orbit that passes through the minimum of the $t_\mathrm{GW} = 10^{-3} t_\mathrm{rlx}$ curve following Eq.\ \eqref{eq:a(e)} (dashed brown line). It can be seen that all orbits that cross into the excluded region while being in the GW-dominated region fall below this evolution line, meaning they will reach the solid green curve, as long as the stochastic kicks due to two-body relaxation are small (see Fig.\ \ref{fig:plane} for the meaning of the other elements). Here we set $M_\bullet = 4 \times 10^6 \, \mathrm{M}_\sun$.}
            \label{fig:reinit}
        \end{figure}

        The reinitialisation procedure becomes biased when the restriction to $t_\mathrm{rlx} < t_\mathrm{max}$ excludes a significant part of the $t_\mathrm{GW} = 10^{-3} t_\mathrm{rlx}$ curve. In this case, it can happen that some orbits well inside the GW emission dominated region but still far from the loss cone, can reach the $t_\mathrm{rlx} = t_\mathrm{max}$ boundary before crossing the $t_\mathrm{GW} = 10^{-3}t_\mathrm{rlx}$ separatrix, thus being erroneously stopped and reinitialised instead of being counted as EMRIs. A clear example of such an orbit is shown by the blue line in Fig.\ \ref{fig:reinit}: the event will clearly produce an EMRI, but according to our reinitialisation criterion is effectively discarded and resimulated.
        \begin{figure*}
            \centering
            \resizebox{\hsize}{!}{\includegraphics{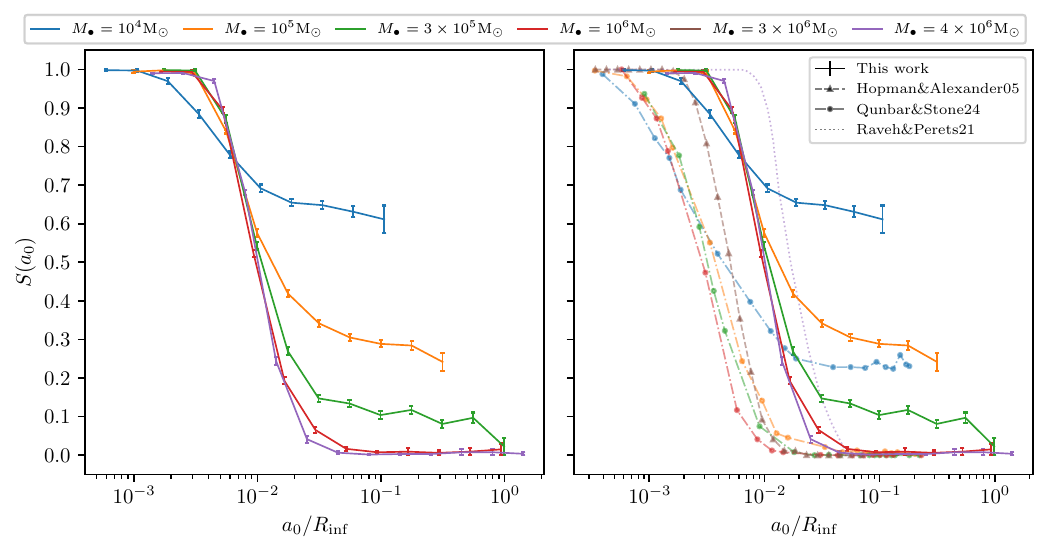}}
            \caption{Function $S(a_0)$ for different $M_\bullet$ scenarios. The bars display 1$\sigma$ uncertainty. The panel on the left displays our results only, while that on the right compares them with those from other works. We note that the other works make different assumptions, both with respect to our work and to each other.}
            \label{fig:res_comp}
        \end{figure*}
        %that are still far from the loss cone butand already dominated by the emission of GWs, therefore preferentially EMRI candidates, are stopped and discarded. 
        As DP candidates are not affected by this issue, the EMRI-to-DP ratio is influenced by these discarded EMRIs. This problem mostly affects the EMRI-to-DP ratio at small initial semi-major axes for large $M_\bullet$, as we show in Sect.\ \ref{sec:main_res}.

        For this reason, we correct our counts of EMRIs to also include orbits which cross into the excluded region while inside the GW-dominated region. Before accepting the orbit as an EMRI, we also verify that the final value of $a$ is below the evolution line of an orbit which evolves as Eq.\ \eqref{eq:a(e)}, where $c_0$ is chosen so that the line crosses the minimum of the $t_\mathrm{GW} = 10^{-3} t_\mathrm{rlx}$ curve (again, see Fig.\ \ref{fig:reinit}).
        This means that, ignoring stochastic kicks due to two-body encounters, the orbit would eventually reach our condition for EMRI formation, while inside the excluded region. We also point out that near the $t_\mathrm{rlx} = t_\mathrm{max}$ curve, the condition $t_\mathrm{GW} < t_\mathrm{rlx}$ is likely sufficient to identify EMRIs anyway, as the stricter condition $t_\mathrm{GW} < 10^{-3} \, t_\mathrm{rlx}$ is mainly introduced to avoid misclassifying DPs as EMRIs, and such a misclassification can only occur for $1-e \lesssim 10^{-4}$, where the $t_\mathrm{GW} = t_\mathrm{rlx}$ curve closely approaches the loss cone.
        %Figure \ref{fig:reinit} shows an example of an orbit where this issue is evident. 

        To summarise, the possible outcomes of a simulation are:
        \begin{enumerate}
            \item a DP, if the separation $r$ between MBH and stellar-mass BH satisfies the condition $r < R_\mathrm{lc}$;
            \item an EMRI, either if \textit{(i)} the condition $t_\mathrm{GW} = 10^{-3} t_\mathrm{rlx}$ is satisfied, or \textit{(ii)} the curve $t_\mathrm{rlx} = t_\mathrm{max}$ is reached with a final $a$ such that the curve $t_\mathrm{GW} = 10^{-3} t_\mathrm{rlx}$ is eventually reached at a later time assuming only GW emission;
            \item a reinitialisation, in all other cases in which the condition $t_\mathrm{rlx} = t_\mathrm{max}$ is reached.
        \end{enumerate}
        Condition \textit{(ii)} for EMRI formation is the correction introduced above, and its importance will be shown in Sect.\ \ref{sec:main_res}.
        Examples of each of these outcomes are shown in Figs.\ \ref{fig:plane}, \ref{fig:cliff}, and \ref{fig:reinit}.
        
        Finally, the code halts if Eq.\ \eqref{eq:delta_vr} has no real solutions when looking for an orbit-averaged kick, or if $\mathcal{E} < 0$ at some point during the simulation (meaning the stellar-mass BH has become unbound). In practice, the former situation excluded a negligible number of simulations, and the latter never occurred.

        For each nuclear cluster configuration, we select a set of possible initial semi-major axes $a_0$ uniformly distributed in logarithm. Then, we adopt the steps described below to randomly draw a different realistic value for the initial eccentricity $e_0$ of each simulation. Given $a_0$, we compute 
        \begin{equation}
            \mathcal{R}_\mathrm{max} = \left.\mathcal{R}\right\rvert_{t_\mathrm{rlx} = t_\mathrm{max}}
        \end{equation}
        and
        \begin{equation}
            \mathcal{R}_\mathrm{min} = \min\left\{\mathcal{R}_\mathrm{lc}, \left.\mathcal{R}\right\rvert_{t_\mathrm{GW} = 10^{-3} t_\mathrm{rlx}}\right\}
        \end{equation}
        at the corresponding energy $\mathcal{E}_0 = G M_\bullet / 2 a_0$. Then, we randomly draw an initial value for $\mathcal{R}_0$ from the probability distribution function
        \begin{equation}
            p(\mathcal{R}) = \frac{\log \left( \mathcal{R}/\mathcal{R}_\mathrm{q} \right)}{\mathcal{R}_\mathrm{max}\left( \ln\left(\mathcal{R}_\mathrm{max}/\mathcal{R}_\mathrm{q}\right)-1\right) - \mathcal{R}_\mathrm{min}\left( \ln\left(\mathcal{R}_\mathrm{min}/\mathcal{R}_\mathrm{q}\right)-1\right)}
        \end{equation}
        to reproduce the Cohn-Kulsrud distribution \citep[][also see Sect.\ \ref{sec:losscone_diff}]{1978ApJ...226.1087C}, where the normalisation factor is such that $\int_{\mathcal{R}_\mathrm{min}}^{\mathcal{R}_\mathrm{max}} \mathrm{d}\mathcal{R}\, p(\mathcal{R}) =1$. Finally, from $\mathcal{R}_0$, we compute
        \begin{equation}
            e_0 = \sqrt{1-\mathcal{R}_0}
        \end{equation}
        and we initialise the orbit with semi-major axis $a_0$ and eccentricity $e_0$ with Newtonian initial conditions, placing the stellar-mass BH at the apocentre.

        In our procedure, we initialise all particles at the apocentre of their initial orbits, rather than assigning a random location along the orbit. We select this point because PN corrections are least significant there and this choice does not bias the likelihood of a particle becoming a DP or an EMRI. Before reaching a disruptive pericentre, the particle will either quickly modify its orbital parameters through two-body scattering (when $t_\mathrm{rlx} \lesssim P$), effectively erasing any memory of the initial condition, or complete more than one period on the same orbit (when $t_\mathrm{rlx} \gtrsim P$), rendering the specific initial location along the orbit irrelevant.

%__________________________________________________________________

\section{Results}\label{sec:results}

    In this section we present the results we obtained running a large number of simulations with the code described in Sect.\ \ref{sec:methods}.   

    \subsection{EMRI-to-DP ratio and cliffhanger EMRIs} \label{sec:main_res}
        \begin{figure*}
            \centering
            \resizebox{\hsize}{!}{\includegraphics{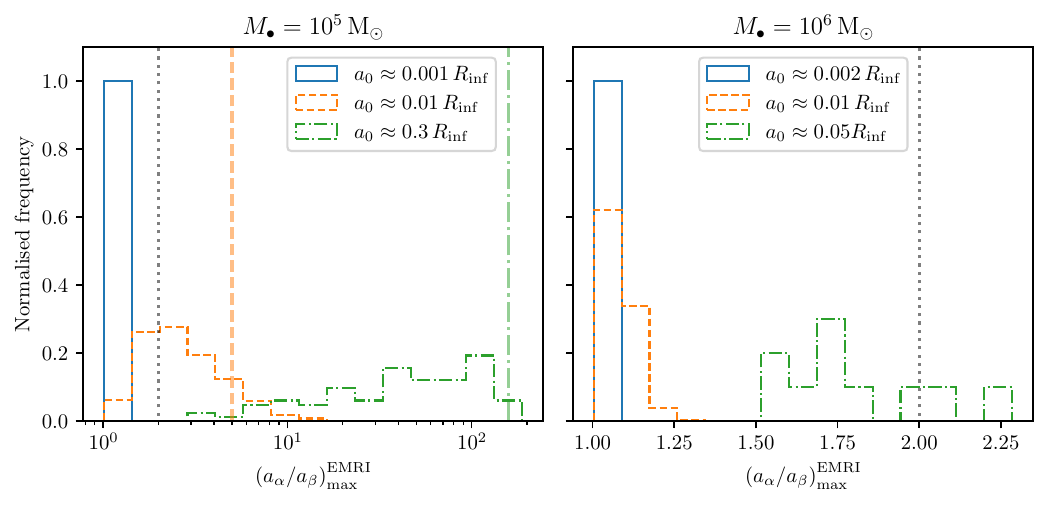}}
            \caption{Distribution of the maximum ratio between $a$ evaluated at subsequent apocentres for EMRIs for selected values of $M_\bullet$ and $a_0$. The black dotted vertical line shows $(a_\alpha/a_\beta)_\mathrm{max}^\mathrm{EMRI} = 2$, which roughly delimits cliffhanger EMRIs to its right. In the left panel, the orange dashed and green dash-dotted vertical lines show $a_0/a_\mathrm{f}^\mathrm{lc}$ from Eq.\ \eqref{eq:a_lc_final}, which is a rough prediction of the highest possible $(a_\alpha/a_\beta)_\mathrm{max}^\mathrm{EMRI}$ ratio. Each histogram is normalised to 1. We note the different $x$-axis scales and binning strategies in the two panels.}
            \label{fig:histo}
        \end{figure*}

        For each simulated central MBH mass, we build the $S(a_0)$ function given the counts of EMRIs and DPs. Given the success/failure nature of the problem (we can see EMRI formation as a success, and DP formation as a failure), we employ the binomial statistic to characterise the uncertainties on our findings. Specifically, we compute the corresponding binomial proportion confidence interval at 1$\sigma$ confidence level using the Wilson score interval \citep{doi:10.1080/01621459.1927.10502953}, an improved version of the commonly used standard Wald interval. The Wilson score interval avoids issues when $S(a_0)$ approaches $0$ or $1$, such as falsely implying certainty or error bars overshooting outside the $[0,1]$ range \citep{10.1214/ss/1009213286}. 

        We report our main findings in the left panel of Fig.\ \ref{fig:res_comp}. For $M_\bullet = 10^6 \, \mathrm{M}_\sun$ and $M_\bullet = 4 \times 10^6 \, \mathrm{M}_\sun$, we recover the expected dichotomy between EMRIs and DPs. In this case, EMRIs can only form from orbits with $a_0 \lesssim a_\mathrm{c}$, while those with $a_0 \gtrsim a_\mathrm{c}$ exclusively result in DPs. We find that the transition between the two regimes occurs over about an order of magnitude in $a_0$ at $a_\mathrm{c} \simeq 10^{-2} R_\mathrm{inf}$, which is consistent within a factor of two with previous works \citep[][QS24]{2005ApJ...629..362H,2021MNRAS.501.5012R}. We believe that the quantitative difference is driven by differences in the nuclear cluster models, in the accuracy of the dynamical model (e.g. PN order) and in the treatment of two-body relaxation. We better explore the impact that some of these factors have on the shape of $S(a_0)$ in Sect.\ \ref{sec:localvsave} and \ref{sec:tests}.

        Conversely, for $M_\bullet < 10^6\, \mathrm{M}_\sun$, we observe a relevant number of EMRIs formed from initially wide orbits, with $a_0 \gg a_\mathrm{c}$. These events are consistent with the cliffhanger EMRI population first seen in Monte Carlo simulations by QS24 and, to the best of our knowledge, never found in previous work.
        %These events are consistent with the novel EMRI population of cliffhanger EMRIs. To the best of our knowledge, these events were first seen in Monte Carlo simulations by QS24, and not observed in any previous work.
        We observe cliffhanger EMRIs forming for $M_\bullet \lesssim 3 \times 10^5 \, \mathrm{M}_\sun$ and they appear to be more common as $M_\bullet$ decreases, consistently with the argument based on the $(1-e,a)$ phase space presented in Sect.\ \ref{sec:cliff_region}. 
        
        Our results for $S(a_0)$ are compared to those obtained by  QS24 and other authors in the right panel of Fig.\ \ref{fig:res_comp}.
        First, we can notice the large scatter in $a_\mathrm{c}/R_\mathrm{inf}$ that spans almost an order of magnitude, from $\approx 3\times10^{-3}$ (QS24) to $\approx 2\times10^{-2}$ \citep{2021MNRAS.501.5012R}. This is likely due to the differences in the parameters of the simulated systems as well as in the treatment of the dynamics and loss cone definition, as we  discuss in Sect.\ \ref{sec:tests}. Focusing on cliffhanger EMRIs,
        QS24 observe their formation only for MBHs with $M_\bullet \lesssim 5 \times 10^4 \,\mathrm{M}_\sun$, a threshold set almost an order of magnitude below our limiting value. Moreover, we see a much larger fraction of EMRIs over DPs at large $a_0$, with a maximum ratio of $S(a_0) \simeq 0.65$ at $a_0 \gg 10^{-2} R_\mathrm{inf}$ for $M_\bullet = 10^4 \,\mathrm{M}_\sun$. As a further difference, our $S(a_0)$ functions hint at a downward trend for $a_0 \gg 10^{-2} R_\mathrm{inf}$, as opposed to tending to a constant at $a_0 \to \infty$. This could be due to the fact that QS24 stopped their investigation at $a_\mathrm{max} \simeq R_\mathrm{inf}/3$, as their implementation does not account for the stellar potential.
        
        Consider now $a_\alpha$ and $a_\beta$ as the values of the semi-major axis of the orbit at two subsequent generic apocentres. For each run that resulted in the formation of an EMRI, we consider the largest ratio $a_\alpha / a_\beta$ to set a criterion to distinguish cliffhanger and traditional EMRIs. Consistently with the reasoning that led us to Eq.\ \eqref{eq:double}, we determine the passage across the cliffhanger region if at some point in the orbital evolution the semi-major axis is cut in half or more during a single period, thus for all the runs where $(a_\alpha/a_\beta)_\mathrm{max}^\mathrm{EMRI} > 2$. This threshold is somehow arbitrary, but offers a quantitative way to identify cliffhanger EMRIs. Figure \ref{fig:histo} shows the result of this investigation for $M_\bullet = 10^5 \, \mathrm{M_\sun}$ and  $M_\bullet = 10^6 \, \mathrm{M_\sun}$. For both MBH masses, all EMRIs forming for $a_0 \ll 10^{-2} R_\mathrm{inf}$ are classical EMRIs. Both cases show an increase in the average value of $(a_\alpha/a_\beta)_\mathrm{max}^\mathrm{EMRI}$ for larger $a_0$, but only for $M_\bullet = 10^5 \, \mathrm{M_\sun}$ the threshold of 2 is crossed. Moreover, we verify that the prediction for the largest possible $(a_\alpha/a_\beta)_\mathrm{max}^\mathrm{EMRI}$ ratio (which occurs at the loss cone edge) given in Eq.\ \eqref{eq:a_lc_final} is verified for $M_\bullet = 10^5 \, \mathrm{M_\sun}$ for large $a_0$, with some deviations for smaller $a_0$. This is likely because the approximation given in Eq.\ \eqref{eq:energy_approx} is less precise for smaller semi-major axes.
        
\begin{figure}
            \centering
            \resizebox{\hsize}{!}{\includegraphics{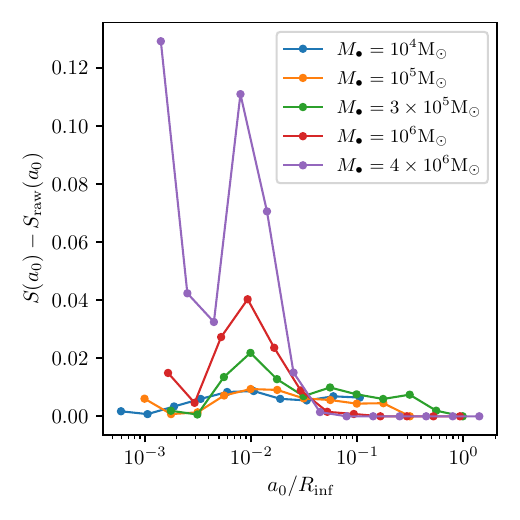}}
            \caption{Impact of the correction to the stopping criterion on the function $S(a_0)$. We show the difference between the function displayed in Fig.\ \ref{fig:res_comp} and $S_\mathrm{raw} (a_0)$, which is the one obtained without correcting the counts of EMRIs.}
            \label{fig:corrections}
        \end{figure}

        Finally, in Fig.\ \ref{fig:corrections}, we show the impact on the function $S(a_0)$ of the correction to the stopping criterion for EMRIs detailed in Sect.\ \ref{sec:stopping}. Largest masses are more affected by the issue, in particular for low $a_0$ and near the inflection point at $a \simeq 10^{-2} R_\mathrm{inf}$. The figure clearly shows that this correction is needed to correctly recover the limit to $1$ of $S(a_0)$ at $a_0 \ll 10^{-2} R_\mathrm{inf}$, while it does not affect the curve at $a_0 \gg R_\mathrm{inf}$ region, where we observe cliffhanger EMRIs.  

        \subsection{Local versus orbit-averaged approach}\label{sec:localvsave}
        
        One of the main objectives of our investigation is that of testing the reliability of the orbit-averaged approximation to describe the complex interactions in nuclear clusters. As explained in Sect.\ \ref{sec:losscone_diff}, this approach is based on the idea of orbit-averaging the local diffusion coefficients, which in reality strongly depend on the exact position $r$ and velocity $v$ of the orbiting body in consideration. In highly eccentric orbits, such as the one needed to form EMRIs, both $r$ and $v$ oscillate by orders of magnitude in a single period, and orbit-averaging diffusion coefficients might lead to a mismodelling of the very different features displayed by relaxation at $r_-$ or $r_+$. Another common practice in the orbit-averaging approach that is likely to lead to inaccuracies is that of accounting for two-body relaxation only once per period to update the orbital elements. This effectively means that once the energy and angular momentum of the particle are updated (usually at apocentre) the orbit is frozen for a full period. This treatment is, however, at odd with the concept of full loss cone itself, as we come across the contradiction of being in the regime where escaping from the loss cone should be easy while being effectively locked onto an orbit, that thus cannot escape.

        The relaxation timescale for a particle with energy $\mathcal{E}$ and pericentre $r_-$ estimates the average time that the particle passes on the orbit before being scattered away. When $r_-$ is much smaller than the semi-major axis of the orbit, it estimates the time before $r_-$ is changed by the order of itself, as energy remains almost constant \citep{1978ApJ...226.1087C}. In this small $r_-$ regime the relaxation timescale for the orbit can be estimated as $t_\mathrm{rlx} \simeq P\,r_-/(q\,r_\mathrm{lc})$ \citep{2013degn.book.....M,2023MNRAS.524.3026B,2024OJAp....7E..48B}, where $P$ is the orbital period.
        Consider now an orbit with energy $\mathcal{E}$ and $r_-<r_\mathrm{lc}$: in a time $P$ it will be reached by a number $\mathrm{d}N$ of stellar objects due to two-body relaxation. In the orbit-averaged approach, they will all be DPs produced in the subsequent orbital period, with a rate from that orbit of $\mathrm{d}\dot{N} = \mathrm{d}N/P$. In the local approach, if $t_\mathrm{rlx}<P$, only the stellar objects that will take less than $t_\mathrm{rlx}$ to reach the pericentre will be DPs, and the corresponding rate will be $\mathrm{d} \dot{N} \simeq \mathrm{d}N \,t_\mathrm{rlx}/P^2 = r_- / (q\,r_\mathrm{lc})\, \mathrm{d}N / P$. Therefore, the orbit-averaged approach will overestimate the rate of DPs for $q\gtrsim 1$ due to the `frozen orbit' approach, as they have $r_-< r_\mathrm{lc}$ by definition. The same reasoning applies to cliffhangers, that are produced at very small pericentres, too. However, being their pericentre larger than $r_\mathrm{lc}$ by definition, the overestimation of cliffhangers in the orbit-averaged assumption will be lower. Consequently, we expect that the orbit-averaged approach will underestimate $S(a)$.
        
\begin{figure}
        \centering
        \resizebox{\hsize}{!}{\includegraphics{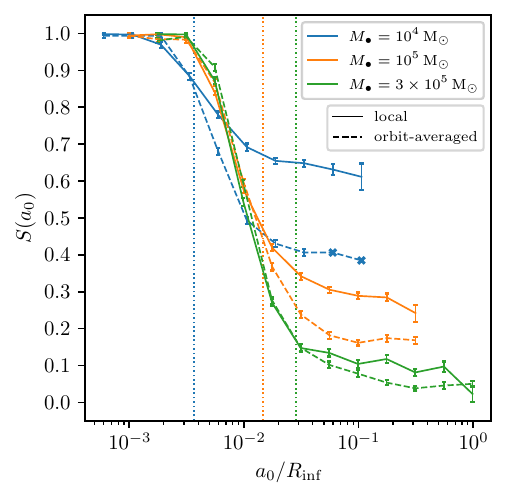}}
        \caption{Comparison of the $S(a_0)$ function obtained by employing the local (solid lines) vs orbit-averaged (dashed lines) treatment of two-body relaxation. The vertical dotted lines mark, for each case, the value of $a_0/R_{\rm inf}$ at which $q=1$, thus marking the transition from the empty (at smaller $a_0$) to the full loss cone regime. The crosses indicate the points for which the orbit-averaged procedure explained in Sect.\ \ref{sec:OA} failed many times.}
        \label{fig:OA}
    \end{figure}

        % This is, in general, justified since in the full loss cone regime, the particle is by definition statistically scattered in and out of the loss cone multiple times in a single orbit. Thus, the probability of being captured right at pericentre, is roughly equal to the probability of the integrated $\Delta J$ along the orbit to put the particle onto a new orbit with $J < J_{\rm LC}$. \lb{Questa frase non mi convince. Sembra in contraddizione con l'idea che si possa scappare dal loss cone: non solo devo finire lì, ma devo starci. In particolare "the probability of being captured right at the pericentre" mi sfugge: si è sempre catturati al pericentro. Potresti chiarire meglio cosa intendi?} \as{No, chiaro, mi sono spiegato male. Luca, riscrivi tu questo paragrafo spiegando per bene perche` ci aspettiamo questo aumento di cliffahanger nella local approx, ok?} However, scattering is a process that continuously modifies the orbital elements along the orbit. Since the cliffhanger region is in $J$ space quite larger than the loss cone, close to pericentre, potentially plunging orbits are likely to cross the cliffhanger region before actually becoming DPs. In other words, we can expect that some DPs estimated with the orbit-averaged approach will indeed statistically be cliffhangers instead.  
    In order to test this expectation we re-ran the three sets of simulations with $M_\bullet = 10^4, 10^5, 3\times10^5 \,\mathrm{M}_\sun$ with the exact same system parameters but employing the orbit-averaged procedure described in Sec.~\ref{sec:OA}. Results of these test runs are compared to our default simulations in Fig.~\ref{fig:OA}, and appear to confirm our expectation. The $S(a_0)$ functions computed in the local vs orbit-averaged simulations nicely match up to the value of $a_0/R_{\rm inf}$ for which $q=1$, which marks the transition from empty to full loss cone (vertical dotted lines in the figure). At larger $a_0$, the orbit-averaged approach tends to produce less cliffhanger EMRIs and more plunges, thus resulting in a lower $S(a_0)$. We note that the trend becomes more severe by lowering the central MBH mass, consistent with the fact that the cliffhanger region becomes larger and larger. In practice, the smaller the central MBH, the bigger gets the cliffhanger region compared to the loss cone, and it becomes increasingly more difficult for the orbit-averaged approximation to capture the detailed dynamics of particle scattered within the cliffhanger region in real time along the orbit close to the pericentre, before getting to the loss cone, thus leading to an over-estimation of DPs.

        %Indeed, by ``freezing'' the shape of the orbit like that, the full loss cone regime cannot be accurately investigated, as we came across the contradiction of being in the regime where escaping from the loss cone should be easy while locked onto an orbit, that thus cannot escape merger......
        
        %In order to test the validity of the orbit-averaged approximation, we perform ....

    \subsection{Detailed investigation of the differences with QS24}   \label{sec:tests}      
        Our main results display some deviations with respect to other works in the literature. In particular the emergence of cliffhanger EMRIs and their abundance does not match findings by QS24. This is not surprising, given the many differences between our simulations and theirs. In particular, we simulate a two-population system, including the effect of solar-mass stars and stellar-mass BHs in the computation of the diffusion coefficient, we adopt a local treatment of two-body relaxation and we evolve the stellar-mass BH using PN equations of motion. Conversely, in QS24 scattering comes from a single population of solar-mass object, is treated as orbit-averaged and the dynamics of the stellar-mass BH is Newtonian. 
        
        In the previous section, we already highlighted the importance of the local treatment of relaxation, which leads to a much larger relative fraction of EMRIs in the full loss cone regime. This can certainly account for some of the differences with QS24. For example in the case of  $M_\bullet = 10^4 \,\mathrm{M}_\sun$, the orbit-averaged tests shown in Fig.~\ref{fig:OA} result in $S(a_0)\approx 0.4$ at large $a_0$, which is much closer to the value of $\approx 0.25$ found in QS24 compared to the $\gtrsim 0.6$ found in our default simulations employing a local treatment of relaxation.
        
\begin{figure}
            \centering
            \resizebox{\hsize}{!}{\includegraphics{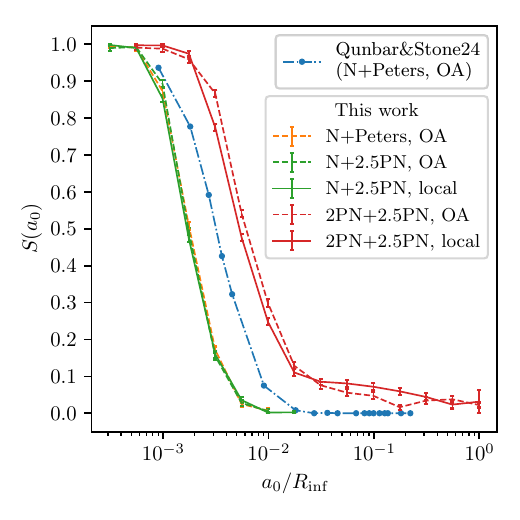}}
            \caption{Function $S(a_0)$ given different assumptions on the dynamics and treatment of two-body relaxation. For the dynamics, we distinguish between Newtonian (N) or PN evolution of the system, and between employing the equations detailed in \citet{1964PhRv..136.1224P} or the 2.5PN term to treat loss of energy and angular momentum due to GW emission. Instead, two-body relaxation can be accounted for using the orbit-averaged approximation (indicated with OA), or locally. In all cases $M_\bullet = 3 \times 10^5\, \mathrm{M}_\sun$ and the MBH is surrounded by a distribution of $m_\star = 1 \,\mathrm{M}_\sun$ stars only.}
            \label{fig:params}
        \end{figure}

        In order to further test how the differences between the QS24 and our studies affect the results, we perform further targeted simulations, focusing on the $M_\bullet = 3 \times 10^5 \,\mathrm{M}_\sun$ case, as it is close to the MBH mass value at which the majority of LISA EMRI detections is expected
        \citep{2017PhRvD..95j3012B}. We try to reproduce the results by QS24\footnote{QS24 do not present results for a $M_\bullet = 3 \times 10^5 \,\mathrm{M}_\sun$ scenario; however, the authors kindly agreed to share the results with us and to let us show them in Fig.\ \ref{fig:params}.} by running our simulations with a setup which is as close as possible to theirs.        
%        \mb{in order} to resemble their \mb{findings} as     close as possible, and then observe how each piece            influences them\mbout{ results by changing it}.
        We use a single population of scattering objects of $m_\star = 1 \,\mathrm{M}_\sun$, distributed according to a Denhen density profile characterised by $M_{\mathrm{tot},\star} = 1425 M_\bullet$, $r_{\mathrm{a},\star} = 223.37 R_\mathrm{inf}$ and $\gamma_\star = 1.75$. With these parameters, our density profile departs from the single $r^{-1.75}$ power law simulated by QS24 by less than 1\% within the MBH influence radius.     
        %to achieve a ratio above $0.99$ up until the influence radius between our numeric density profile, defined in Eq.\ \eqref{eq:n}, and the one assumed in QS24. Figure \ref{fig:ratio_n} shows this ratio as a function of $r$. We also remove the stellar-mass BH population to match their work.
%        We use $m_\star = 1 \mathrm{M}_\sun$, $M_{\mathrm{tot},\star} = 1425 M_\bullet$, $r_{\mathrm{a},\star} = 223.37 R_\mathrm{inf}$ and $\gamma_\star = 1.75$ to achieve a ratio above $0.99$ up until the influence radius between our numeric density profile, defined in Eq.\ \eqref{eq:n}, and the one assumed in QS24. Figure \ref{fig:ratio_n} shows this ratio as a function of $r$. We also remove the stellar-mass BH population to match their work.

        \citet{2024PhRvL.133n1401Q} use Newtonian dynamics, and take into account emission of GWs through the equations detailed in \citet{1964PhRv..136.1224P}. We do the same by setting $H_1 = H_2 = H_{2.5} = 0$ and adding the contributions
        \begin{equation}
            \Delta \mathcal{E}_\mathrm{GW} = P \left\langle \frac{\mathrm{d}\mathcal{E}}{\mathrm{d}t } \right\rangle_\mathrm{GW} \, , \qquad \Delta J_\mathrm{GW} = P \left\langle \frac{\mathrm{d}J}{\mathrm{d}t } \right\rangle_\mathrm{GW}
        \end{equation}
        to the orbital sums of the stochastic kicks $\Delta \mathcal{E} = \sum_i \delta \mathcal{E}_i$ and $\Delta J = \sum_i \delta J_i$ once per orbit, before applying the velocity kick following the orbit-averaged procedure described in Sect.\ \ref{sec:OA}.

        We report the results of these tests in Fig.\ \ref{fig:params}.
        Despite our attempt to reproduce the system employed in QS24 as close as possible, we cannot replicate their results. Although the shape of the $S(a_0)$ function is very similar, our $a_\mathrm{c}$ is approximately a factor of two smaller than what they found. It is likely that this discrepancy stems from the neglect of the stellar potential in QS24, or in differences in the specific implementation of orbit-averaging. Further investigation is still needed to get a definitive answer. Moreover, we observe that $S(a_0)$ is not modified by using the 2.5PN term to account for GW radiation instead of the equations detailed in \citet{1964PhRv..136.1224P}. As discussed in the previous section, for this central MBH mass, the local treatment of two-body relaxation does not have a strong impact on the $S(a_0)$ function. 
        We find that in Newtonian dynamics the local treatment results in no discernible deviation of $S(a_0)$ compared to the orbit-averaged approximation. At second PN order instead, we observe some more significant deviations. In particular, as noted in the previous section, our local approach predicts only slightly more DPs for $a_0 \simeq 3\times 10^{-3} R_\mathrm{inf}$ and more EMRIs for $a_0 \gtrsim 2\times10^{-2} R_\mathrm{inf}$.
        Finally, a large difference comes from the choice of dynamics. When going from Newtonian to PN dynamics, $S(a_0)$ is shifted to larger semi-major axes and cliffhanger EMRIs become more likely. This is because PN orbits can get closer to the MBH without being accreted as DPs ($R_{\rm lc}=5.6R_\mathrm{g}$ instead of $8R_\mathrm{g}$), thus having the chance of experiencing much larger jumps in the $(a, 1-e)$ plane along the loss cone line. This, in turn, means that cliffhanger EMRIs can form from larger $a_0$, consistent with Fig.\ \ref{fig:params}. 
        
        These results show that $a_\mathrm{c}$ (i.e.\ the EMRI-DP separatrix) as well as the abundance of cliffhanger EMRIs depends on the treatment of the stellar-mass BH orbit dynamics. Since PN orbits are a better approximation to GR than Newtonian gravity, we speculate that our PN results are closer to reality than those obtained by employing Newtonian dynamics. However, a full geodesic orbital description in GR would be needed to provide an accurate characterisation of the $S(a_0)$ function.

    \subsection{Effect on EMRI rates} \label{sec:rates}
%    \textcolor{red}{DA MODIFICARE PER AGGIUNGERE I NUOVI CASI M=1e4, M=1e5}
    
    We now apply our results on $S(a_0)$ to assess the impact of cliffhanger EMRIs on the rates of EMRI predicted by Fokker-Planck codes. Classically, the rate of EMRIs is computed from the differential rate in energy of loss cone captures
    \begin{equation}
        \mathcal{F}_\mathrm{lc}(\mathcal{E}) \equiv \frac{\mathrm{d} \dot{N}}{\mathrm{d}\mathcal{E}}
    \end{equation}
    that can be directly computed during the evolution \citep{1978ApJ...226.1087C}. The total rate of EMRIs and DPs per unit time can be computed as follows:
    \begin{equation}
        \dot{N}_\mathrm{tot} = \int_{0}^{\infty} \mathrm{d}\mathcal{E} \, \mathcal{F}_\mathrm{lc}(\mathcal{E})\,.
    \end{equation}
    In general, $S(a_0)$ acts on the differential rate as a transfer function, giving the fraction of loss cone events happening at semimajor axis $a_0$ that will result in EMRIs. The function $1-S(a_0)$ gives, symmetrically, the fraction of DPs.
    In order to apply the transfer function $S(a_0)$, one needs to map the orbit in a generic potential to a typical scale radius, interpreted as the semi-major axis when the stellar object enters the loss cone or the GW-dominated region, meaning it is considered to be captured. A reliable choice for objects captured at a given energy is the radius $r_\mathrm{c}$ of the circular orbit with the same energy, that reduces to the semi-major axis in the Keplerian regime:
    \begin{equation}
        S(\mathcal{E}) = S(r_\mathrm{c}(\mathcal{E})) \, .
    \end{equation}
    Another possibility is $a_0 = (r_- + r_+)/2$, but it is affected by the stellar potential at smaller $\mathcal{E}$. The two mappings for typical loss cone orbits depend on the total potential evaluated at $r_\mathrm{c} = a_0$ and $r_+ \simeq 2\,a_0$ respectively.
    The rates of DPs and EMRIs can be computed as
    \begin{align}\label{eq:newLCrates}
        \dot{N}_\mathrm{DP} &= \int_0^\infty \mathrm{d}\mathcal{E} \; \left[ 1 - S(\mathcal{E})\right]\; \mathcal{F}_\mathrm{lc}(\mathcal{E})\,,\\
        \dot{N}_\mathrm{EMRI} &= \int_0^\infty \mathrm{d}\mathcal{E} \; S(\mathcal{E}) \; \mathcal{F}_\mathrm{lc}(\mathcal{E}) \, .
    \end{align}
    When using the classical discriminant between EMRIs and DPs, one sets $S(\mathcal{E})  = \Theta(\mathcal{E} - \mathcal{E}_\mathrm{c})$, where $\Theta(x)$ is the Heaviside step function, that is zero when its argument is negative and one otherwise, and $\mathcal{E}_\mathrm{c}$ is the energy such that $r_\mathrm{c}(\mathcal{E}_\mathrm{c})=a_\mathrm{c}$. Accordingly, the classical rate of the two classes of captures, which we denote with the upper index `cl', is
    \begin{align}\label{eq:oldLCrates}
        \dot{N}_\mathrm{DP}^\mathrm{cl} &= \int_0^{\mathcal{E}_\mathrm{c}} \mathrm{d}\mathcal{E} \,\mathcal{F}_\mathrm{lc}(\mathcal{E})\, ,\\
        \dot{N}_\mathrm{EMRI}^\mathrm{cl} &= \int_{\mathcal{E}_\mathrm{c}}^{\infty} \mathrm{d}\mathcal{E} \,\mathcal{F}_\mathrm{lc}(\mathcal{E}) \, .
    \end{align}
    
\subsubsection{The case $M_\bullet=3\times10^5\, {\rm M}_\odot$: implications for the most relevant LISA systems}
    
    We first specialise to the case $M_\bullet = 3\times10^5\, {\rm M}_\odot$, which is where the bulk of LISA detections are expected, and use the code presented in \citet{2022MNRAS.514.3270B} to simulate the relaxation of the nuclear cluster surrounding it. The initial Dehnen profile of stars and stellar-mass BHs has $\gamma_\star=\gamma_\bullet=1.5$, while the other properties match those of Sect.~\ref{sec:setup}. At later times, the system relaxes and two-body scattering in the inner region are dominated by stellar-mass BHs, whose distribution reaches $\gamma_\bullet=1.7$, as in Sect.~\ref{sec:setup}. The code accounts for a non-evolving stellar potential set by the initial conditions and includes the effects of two-body relaxation, but does not account for GWs emission. Since the GWs driven region of phase space encloses the loss cone for $a < a_\mathrm{c}$, we interpret $\mathcal{F}_\mathrm{lc}$ for $\mathcal{E}>\mathcal{E}_\mathrm{c}$ as the differential rate entering the GW dominated region.\footnote{Some recent works refer to the line $t_\mathrm{GW}\simeq t_\mathrm{rlx}$ as the loss cone for $a < a_\mathrm{c}$ \citep{2024PhRvD.109d3005B, 2024arXiv240607627K, 2024ApJ...977....7R}. In this work, with `loss cone curve' we always refer to the $r_- = R_\mathrm{lc}$ curve.}
    For this system $R_\mathrm{inf} = 0.57$ pc and $\mathcal{E}_\mathrm{c} = 63 \, \sigma^2_\mathrm{inf}$, where $\sigma_\mathrm{inf} = \sigma (R_\mathrm{inf})$. We build $S(\mathcal{E})$ by interpolating the function shown in Fig.~\ref{fig:res_comp} linearly in the $\log$-$\log$ space. We extrapolate as $S(\mathcal{E}) = 0$ for $\mathcal{E}\to0$ ($a_0 \to \infty$), and $S(\mathcal{E}) = 1$ for $\mathcal{E}\to\infty$ ($a_0 \to 0$).
    
\begin{figure}
        \centering
        \resizebox{\hsize}{!}{\includegraphics{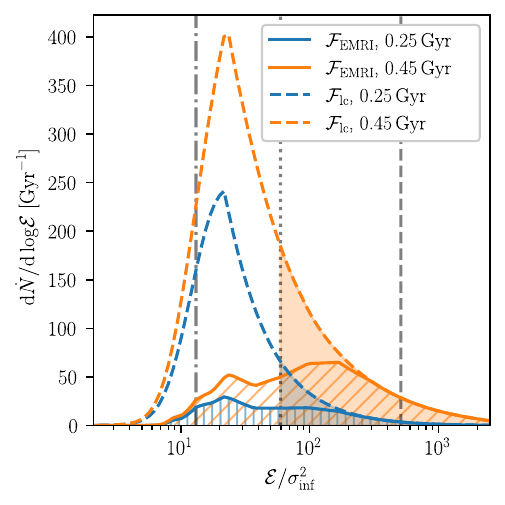}}
        \caption{Differential rate of EMRIs $\mathcal{F}_\mathrm{EMRI}$ (solid lines) and of EMRIs and DPs $\mathcal{F}_\mathrm{lc}$ (dashed lines) as functions of energy. Quantities refer to a nuclear cluster with $M_\bullet = 3\times10^5\, \mathrm{M_\sun}$ that has relaxed for 25 Myr (blue lines) and 45 Myr (orange lines). We report for reference vertical lines at the values of $\mathcal{E}$ such that $r_\mathrm{c}(\mathcal{E})$ is $0.1\,R_\mathrm{inf}$ (dash-dotted), $0.01\, R_\mathrm{inf}$ (dotted) and $0.001\,R_\mathrm{inf}$ (dashed). The colour-filled   area of $\mathcal{F}_\mathrm{lc}$ gives the classical EMRI rate at $r_\mathrm{c}(\mathcal{E}) > 0.01\,R_\mathrm{inf}$, to be compared with the hatched area of $\mathcal{F}_\mathrm{EMRI}$ over the full energy range. Apparent areas in the plot are proportional to the various integrals of the form $\int d\mathcal{E} \,\mathcal{F}$.}
        \label{fig:rates}
    \end{figure}

    In Fig.~\ref{fig:rates}, we show the differential rate $\mathcal{F}_\mathrm{lc}$ computed at $t=25$ Myr and $t=45$ Myr, both after the initial numerical transient (coming from the perfectly isotropic initial conditions), but before the peak of the classical EMRI rates, which happens at $t=75$ Myr. We also show the differential rate of EMRIs
    \begin{equation}
        \mathcal{F}_\mathrm{EMRI}(\mathcal{E}) \equiv \mathcal{F}_\mathrm{lc}(\mathcal{E}) \, S(\mathcal{E})
    \end{equation}
    for a direct comparison with its classical estimate
    \begin{equation}
        \mathcal{F}^\mathrm{cl}_\mathrm{EMRI}(\mathcal{E}) \equiv \mathcal{F}_\mathrm{lc}(\mathcal{E}) \, \Theta(\mathcal{E}-\mathcal{E}_\mathrm{c})\,.
    \end{equation}
    
    The instantaneous rate of EMRIs and DPs according to our and the classical prescriptions are reported in Table \ref{tab:LCrates}. At $t=25$ Myr, when EMRIs classically account for $15\%$ of the total loss cone event rate, Eq.~\eqref{eq:newLCrates} gives a total rate of EMRIs larger by ${\sim}35\%$. The rate of DPs is only reduced by ${\sim}5\%$, as they are overall more numerous. Therefore, with our new estimates EMRIs account for $20\%$ of the total loss cone event rate. At $t=45$ Myr, EMRIs classically account for ${\sim}30\%$ of the total loss cone event rate, with our new estimate of the EMRI and DP rates comparable to the classical estimates.
    \begin{table*}[]
        \caption{\label{tab:LCrates}Instantaneous EMRI+DP ($\dot{N}$), EMRI ($\dot{N}_\mathrm{EMRI}$), and DP ($\dot{N}_\mathrm{DP}$) rates produced in the snapshots we investigated.}
        \centering
        \begin{tabular}{cccccccc}
        $M_\bullet$ [$\mathrm{M}_\sun$] & $t$ [Myr] & $\dot{N}$ [yr$^{-1}$] & $\dot{N}^\mathrm{cl}_\mathrm{EMRI}$  [yr$^{-1}$]& $\dot{N}_\mathrm{EMRI}$  [yr$^{-1}$] & $\dot{N}^\mathrm{cl}_\mathrm{DP}$  [yr$^{-1}$]& $\dot{N}_\mathrm{DP}$  [yr$^{-1}$]\\ \midrule
        $10^4$ & $0.58$ & $5.9\times10^{-7}$ & $5.3\times10^{-7}$ & $4.0\times10^{-7}$ & $6.5\times10^{-8}$ & $1.9\times 10^{-7}$\\        
        $10^5$ & $11$ & $6.6\times10^{-7}$ & $3.1\times10^{-7}$ & $2.9\times10^{-7}$ & $3.7\times10^{-7}$ & $3.8\times 10^{-7}$\\
        $3 \times 10^5$ & $25$ & $3.5\times10^{-7}$ & $4.9\times10^{-8}$ & $6.7\times10^{-8}$ & $3.0\times10^{-7}$ & $2.8\times 10^{-7}$\\
        $3 \times 10^5$ & $45$ & $7.4\times10^{-7}$ & $2.0\times10^{-7}$ & $2.1\times10^{-7}$ & $5.4\times10^{-7}$ & $5.3\times 10^{-7}$\\
        \end{tabular}
        \tablefoot{We report estimates according to our formulation Eq.~\eqref{eq:newLCrates} (no superscript) based on the transfer function $S(a_0)$, and the classical estimates Eq.~\eqref{eq:oldLCrates} (superscript $\mathrm{cl}$) based solely on the critical semi-major axis $a_\mathrm{c} = 0.01\, R_\mathrm{inf}$.}
    \end{table*}
        \begin{figure*}
            \centering
        \resizebox{\hsize}{!}{\includegraphics{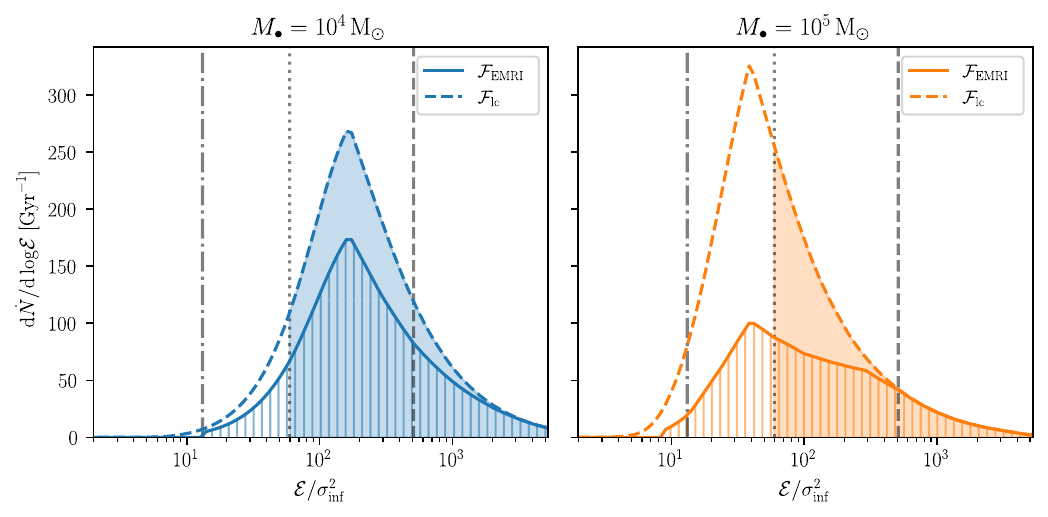}}
        \caption{Differential rate of EMRIs $\mathcal{F}_\mathrm{EMRI}$ (solid lines) and of EMRIs and DPs $\mathcal{F}_\mathrm{lc}$ (dashed lines) as functions of energy. The quantities refer to a nuclear cluster with $M_\bullet = 10^4\, \mathrm{M_\sun}$ that has relaxed for 580 kyr (left panel) and one with $M_\bullet = 10^5\, \mathrm{M_\sun}$ that has relaxed for 11 Myr (right panel). We report for reference vertical lines at the values of $\mathcal{E}$ such that $r_\mathrm{c}(\mathcal{E})$ is $0.1\,R_\mathrm{inf}$ (dash-dotted), $0.01\, R_\mathrm{inf}$ (dotted) and $0.001\,R_\mathrm{inf}$ (dashed). The colour-filled  area of $\mathcal{F}_\mathrm{lc}$ gives the classical EMRI rate at $r_\mathrm{c}(\mathcal{E}) > 0.01\,R_\mathrm{inf}$, to be compared with the hatched area of $\mathcal{F}_\mathrm{EMRI}$ over the full energy range. Apparent areas in the plot are proportional to the various integrals of the form $\int d\mathcal{E} \,\mathcal{F}$.}
        \label{fig:rates2}
    \end{figure*}
    
    Our model predicts a fraction of EMRIs produced in the classical region $\mathcal{E} > \mathcal{E}_\mathrm{c}$ of $45\%$ at $t=25$ Myr, and $60\%$ at $t=45$ Myr. Even with a small value $S(a_0)\simeq 0.1$ at large semi-major axes, cliffhanger EMRIs are a relevant contribution to the overall count of EMRIs due to the shape of the loss cone flux.

    Aside from the rates, the inclusion of $S(a_0)$ has three direct consequences on the energy distribution of EMRIs and DPs:
    \begin{itemize}
        \item about $10 \%$ of DPs are produced for $a_0 < a_\mathrm{c}$, mainly between $\mathcal{E}_\mathrm{c}$ ($r_\mathrm{c} = 0.01 \, R_\mathrm{inf}$) and $5\, \mathcal{E}_\mathrm{c}$ ($r_\mathrm{c} = 0.002 \, R_\mathrm{inf}$);
        \item there is a significant deviation from the classical expectation that most EMRIs are produced with $a_0 \simeq a_\mathrm{c}$, even restricting to $a_0 \leq a_\mathrm{c}$;
        \item the differential rate of EMRIs has a local maximum where the overall loss cone rate has a peak, in this case at $r_\mathrm{c} \simeq 0.07 R_\mathrm{inf}$, corresponding to cliffhangers.
    \end{itemize}
    The first two points depend mainly on the fact that $S(a_0)$ takes approximately an order of magnitude in semi-major axis (or, equivalently, in energy) to smoothly decrease from 1 to its minimum value $S_\mathrm{min}$, with $S_\mathrm{min} < S(a_\mathrm{c}) < 1$ at any value of $M_\bullet$. This shows the limitations of approximating $S(\mathcal{E})$ as a step function. The last point, on the other hand, is exclusively caused by the novelty that $S(a_0) > 0$ even for $a_0 \gg a_\mathrm{c}$ introduced by QS24.

\subsubsection{Effects on EMRIs from lower central MBH masses}

    Having explored in detail the most relevant LISA  case, it is also interesting to have a look at the EMRI rate implications for lower MBH masses, where cliffhanger EMRIs have been shown to dominate the $S(a_0)$ function. Results are shown in Fig.~\ref{fig:rates2} and quantified in Table~\ref{tab:LCrates} for the cases $M_\bullet = 10^4, 10^5\, {\rm M}_\odot$. Rates are computed at $t/t_\mathrm{p}=0.6$, where $t_\mathrm{p}$ is the peaking time of the classical EMRI rate for the system under scrutiny \citep{2022MNRAS.514.3270B}.
    
    Despite cliffhangers becoming increasingly more common at lower MBH masses, their effect on the overall EMRI rates gets progressively less important. This is mainly because the peak energy of the loss cone flux, $\mathcal{E}_{\rm lc,p}$, becomes comparable (and even bigger) then $\mathcal{E}_\mathrm{c}$ at low MBH masses. In the $M_\bullet = 10^5\, {\rm M}_\odot$ case, $\mathcal{E}_\mathrm{c}\gtrsim \mathcal{E}_{\rm lc,p}$, therefore adding a significant population of cliffhanger EMRIs at lower energies has a relevant impact on the overall EMRI population. Despite the overall EMRI rate being substantially unaffected (Table~\ref{tab:LCrates}) we find that ${\sim}37 \%$ of those EMRIs are cliffhanger and that the peak of the EMRI flux occurs at $a>a_\mathrm{c}$ ($\mathcal{E}<\mathcal{E}_\mathrm{c}$) which might result in extremely eccentric systems. Conversely, the overall EMRI rate decreases in the $M_\bullet = 10^4\, {\rm M}_\odot$ case, with cliffhangers accounting only to ${\sim}10 \%$ of the overall EMRI population. Notably, for such a low MBH mass, $\mathcal{E}_\mathrm{c} < \mathcal{E}_{\rm lc,p}$ meaning that most EMRIs form well inside $a_\mathrm{c}$ potentially resulting in lower eccentric systems.

    This analysis shows that while the impact of the detailed treatment of nuclear dynamics and the appearance of cliffhangers on the EMRI rate might be only moderate, the energy distribution of EMRIs (and hence their detailed properties) can be highly affected. In particular, a simplistic treatment of EMRIs as systems mostly forming at $a=a_\mathrm{c}$ fails in capturing the rich and MBH mass dependent features of this class of GW sources.
        
%__________________________________________________________________

\section{Discussion and conclusions} \label{sec:disc_conc}

    In this work we developed a formalism to locally account for the effects of two-body relaxation on the formation of EMRIs and DPs around an MBH at the core of a nuclear cluster, for which we considered a two-population model. We examined a binary system comprised of an MBH and a stellar-mass BH orbiting around it, evolving the system by using PN dynamics up to the 2.5PN term. We ran a number of simulations and studied the EMRI-to-DP ratio as a function of the initial semi-major axis $a_0$ of the orbit through the examination of the function $S(a_0)$, for five different central MBH masses. Our main findings can be summarised as follows:
    \begin{enumerate}
        \item For low MBH masses the classical picture, for which there is a critical value $a_\mathrm{c}$ such that EMRIs only form at $a_0 \ll a_\mathrm{c}$ while no EMRIs occur at $a_0 \gg a_\mathrm{c}$, breaks down. Instead, we find a significant population of EMRIs forming for large values of $a_0$. Such a  population is consistent with the novel phenomenon of cliffhanger EMRIs, first identified in QS24.
        \item  Cliffhanger EMRIs are more common for low-mass MBHs because there is a larger region in the $(1-e,a)$ phase space which allows for a drastic reduction of the orbit in a single pericentre passage due to GW radiation. Following this analytical argument, we predict that the maximum MBH mass expected to produce cliffhanger EMRIs, given our assumption, is roughly $3 \times 10^5 \, \mathrm{M_\sun}$, which was verified in our numerical investigation.
        \item Employing PN dynamics to evolve the system leads to more EMRIs, strongly modifying the shape of the function $S(a_0)$ by shifting $a_\mathrm{c}$ to larger values and boosting the tail of cliffhanger EMRIs produced at $a_0 \gg a_\mathrm{c}$.
        \item The choice of locally accounting for two-body relaxation does significantly influence the EMRI-to-DP ratio for $M_\bullet \lesssim 10^5 \, \mathrm{M_\sun}$, leading to a much larger fraction of cliffhanger EMRIs compared to that found in the orbit-averaged approximation.
        \item Cliffhanger EMRIs account for a significant fraction of the total EMRI rate for the MBH masses where the loss cone capture differential rate peaks at $a_\mathrm{p} > a_\mathrm{c}$ (up to $55\%$ in the snapshots we tested). This is due to a large fraction of EMRIs being produced near this peak.
    \end{enumerate}

    These results might have a significant impact on the estimates of EMRI detection rates by LISA. Currently, these estimates are still highly uncertain, ranging from a few to a few thousand EMRIs detected per year \citep{2017PhRvD..95j3012B}, and do not account for the novel population of cliffhanger EMRIs. Furthermore, we note that many cliffhanger EMRIs get to the GW emission region while grazing the loss cone edge, which might imply a significant sub-population of extremely eccentric EMRIs in the LISA band. We defer the investigation of observable cliffhanger EMRI properties by LISA to future work.

    Interestingly, we observed that our results are quite sensitive to the choice of the loss cone radius $R_\mathrm{lc}$.  As explained in Sect. \ref{sec:tests}, part of the reason for the differences between the $S(a_0)$ obtained with Newtonian and PN dynamics likely stems from switching from $R_\mathrm{lc} = 8 R_\mathrm{g}$ to $R_\mathrm{lc} = 5.6 R_\mathrm{g}$, as plunges are more difficult to form and the cliffhanger region grows. For this reason, we believe that the closest approach to a BH on an eccentric orbit should be better explored at PN order, similarly to how it has  already been done for circular orbits \citep[e.g.][]{2024LRR....27....2S}. Compared to Schwarzschild MBHs, spinning MBHs present innermost stable orbits which are closer or further away, respectively, depending on whether the orbit is prograde or retrograde. This again influences the loss cone radius, and consequently the EMRI-to-DP ratio. For this reason, we plan on updating our code to introduce spin effects.  

    Despite improving on realism by not employing orbit-averages in accounting for two-body relaxation, our approach is still limited on some fronts. Our analysis currently addresses spherical systems with a central Schwarzshild MBH. However, these results might not hold for axisymmetric or triaxial systems due to aspherical two-body relaxation and orbits naturally penetrating the loss cone \citep{2013ApJ...774...87V,2024arXiv240518500K}, or for spinning MBHs \citep{2013MNRAS.429.3155A}. Furthermore, the two-component Bahcall-Wolf profile we assumed might underestimate the number of stellar-mass BHs for $r < R_\mathrm{inf}$, as previous works \citep{2006ApJ...645L.133H,2022MNRAS.514.3270B,2024ApJ...977....7R} have found more segregated profiles. As a result, in our simulations stars dominate two-body relaxation over the entire $r$ range, while for a completely relaxed system stellar-mass BHs should dominate for $r \lesssim 0.1 R_\mathrm{inf}$. While this could affect the shape of $S(a_0)$, and in particular the value of $a_\mathrm{c}$, we note  that for the EMRI rate computations of Sect.\ \ref{sec:rates} the density profiles were consistently relaxed using the code presented in \citet{2022MNRAS.514.3270B}, thus resulting in a more segregated distribution of stellar-mass BHs.
    
    Finally, even restricting the discussion to the class of systems we are investigating, we explored a limited region of the $(1-e,a)$ plane. Since this choice was not biased (at least for low MBH masses) towards EMRIs or DPs, this did not influence our results on the EMRI-to-DP ratio, but it required the introduction of corrections on the counts of EMRIs for high MBH masses and limits the maximum $a_0$ that we can explore. This is particularly true for low MBH masses, for which we are not able to reach $a_0 \approx R_\mathrm{inf}$, as most simulated orbits starting from large $a_0$ result in reinitialisations, thus requiring a very large number of runs to find the EMRI-to-DP ratio within a reasonably small error. We plan on solving this issue in a follow-up work, as pushing the exploration to larger $a_0$ values is fundamental in order to understand whether cliffhanger EMRIs allow $S(a_0)$ to reach a plateau $S(\infty) \neq 0 $ or not.

% \as{Perhaps we can add: "Finally, we note that many cliffhangr EMRIs get to the GW emission region while 'grazing' the loss cone separatrix, which might imply a significant sub-population of extremely eccentric EMRIs in the LISA band, although we defer the investigation of observable cliffhanger EMRI properties to future work."} \dm{Credo abbia più senso metterlo nelle discussioni, ho provato a inserirlo.}    
%__________________________________________________________________

\begin{acknowledgements}
    We thank Ismail Qunbar and Nicholas C.\ Stone for useful discussions and willingness to share data with us. \\
    DM acknowledges that this publication was produced while attending the PhD program in Space Science and Technology at the University of Trento, Cycle XXXIX, with the support of a scholarship financed by the Ministerial Decree no.\ 118 of 2nd March 2023, based on the NRRP - funded by the European Union - NextGenerationEU - Mission 4 ``Education and Research'', Component 1 ``Enhancement of the offer of educational services: from nurseries to universities'' - Investment 4.1 ``Extension of the number of research doctorates and innovative doctorates for public administration and cultural heritage'' - CUP E66E23000110001. MB acknowledges support provided by MUR under grant ``PNRR - Missione 4 Istruzione e Ricerca - Componente 2 Dalla Ricerca all'Impresa - Investimento 1.2 Finanziamento di progetti presentati da giovani ricercatori ID:SOE\_0163'' and by University of Milano-Bicocca under grant ``2022-NAZ-0482/B''. AS acknowledges financial support provided under the European Union’s H2020 ERC Consolidator Grant ``Binary Massive Black Hole Astrophysics'' (B Massive, Grant Agreement: 818691).
\end{acknowledgements}
\newpage

%%%%%%%%%%%%%%%%%%%%%%%%%%%%%%%%%%%%%%%%%%%%%%%%%%%%%%%%%%%%%%%%%%%%%%%%%%%%
%%%%%%%%%%%%%%%%%%%%%%%%%%%%%%%%%%%%%%%%%%%%%%%%%%%%%%%%%%%%%%%%%%%%%%%%%%%%

\bibliographystyle{aa} % style aa.bst
\bibliography{bibliography.bib} % your references Yourfile.bib

\begin{appendix}

\section{Diffusion coefficients}\label{app:diff_coe}

    In this work, we consider a stellar-mass BH of mass $m_\bullet$ and velocity magnitude $v$ having a sequence of two-body encounters with background bodies, grouped into populations (identified by the index $\alpha$) based on their masses $m_\alpha$. We decompose its velocity change as the sum of $\Delta v_\parallel$, parallel to the initial line of motion, and $\Delta v_\perp$, in the plane orthogonal to the original direction of motion.
    
    We use the notation $\mathrm{D}[X]$ to indicate the expectation value of the generic quantity $X$ per unit of time. The diffusion coefficients we use in this work are \citep{1943ApJ....97..255C,2008gady.book.....B,2013degn.book.....M}:
    \begin{align}
        \mathrm{D}\left[\Delta v_\parallel\right] &= -16\pi^2 G^2 \ln \Lambda \sum_\alpha m_\alpha (m_\bullet+m_\alpha) \, \mathcal{I}_{2,\alpha}(r,v) \, ,\\
        \mathrm{D}\left[\Delta v_\perp\right] &= 0 \, ,\\
        \mathrm{D}\left[\left(\Delta v_\parallel\right)^2\right] &= \frac{32 \pi^2}{3} G^2 \ln\Lambda \, v \sum_\alpha m_\alpha^2 \left( \mathcal{I}_{4,\alpha} (r,v) + \mathcal{J}_{1,\alpha} (r,v) \right) \\
        \mathrm{D}\left[\left(\Delta v_\perp\right)^2\right] &=
        \begin{aligned}[t]
            &\frac{32 \pi^2}{3} G^2 \ln \Lambda \, v \, \times \\
            & \times \sum_\alpha m_\alpha^2 \left( 3\mathcal{I}_{2,\alpha} (r,v) -\mathcal{I}_{4,\alpha} (r,v) + 2\mathcal{J}_{1,\alpha} (r,v) \right) \, .
        \end{aligned}
    \end{align}
    Here $\ln{\Lambda}$ is the Coulomb logarithm, which we set to $10$ as an order of magnitude approximation \citep{2003MNRAS.344...22S,2013CQGra..30x4005M}, while the functions $\mathcal{I}_{k,\alpha}$ and $\mathcal{J}_{k,\alpha}$ are defined as
    \begin{align}
        \mathcal{I}_{k,\alpha} (r,v) &= \int^v_0 \mathrm{d}v' \left(\frac{v'}{v}\right)^k f_\alpha(r,v') \, , \\
        \mathcal{J}_{k,\alpha} (r,v) &= \int^\infty_v \mathrm{d}v' \left(\frac{v'}{v}\right)^k f_\alpha(r,v') \, ,
    \end{align}
    with $f_\alpha$ being the distribution function of objects in population $\alpha$ on the six-dimensional phase-space of position and velocity. Assuming that the distribution of scatterers is spherical, and approximating it as locally isotropic in velocity, $f_\alpha$ depends only on the magnitudes of its arguments $r$ and $v$.

    In practice, the diffusion coefficients tell us how the velocity of the stellar-mass BH changes per unit time as it moves with velocity $v$ through the distributions of background bodies at distance $r$ from their centres. Notably, the perpendicular variation of the velocity $\mathrm{D}\left[\Delta v_\perp\right]$ vanishes since the stellar-mass BH is equally likely to be deflected along any of the infinite directions perpendicular to its motion, while the parallel variation $\mathrm{D}\left[\Delta v_\parallel\right] < 0$ describes the effect of dynamical friction \citep{1943ApJ....97..255C}.

    At Newtonian order, $v = \sqrt{2 (\psi - \mathcal{E})}$ for bound orbits. Under this approximation, we can express $\mathcal{I}_{k,\alpha}$ and $\mathcal{J}_{k,\alpha}$ as functions of $\psi(r)$ and $\mathcal{E}(r,v)$:
    \begin{align}
        \mathcal{I}_{k,\alpha} (\psi, \mathcal{E}) &= \frac{1}{\sqrt{2(\psi-\mathcal{E})}} \int_\mathcal{E}^\psi \mathrm{d}\mathcal{E'} \left(\frac{\psi-\mathcal{E'}}{\psi-\mathcal{E}}\right)^{(k-1)/2} f_\alpha (\mathcal{E'}) \, , \\
        \mathcal{J}_{k,\alpha} (\psi, \mathcal{E}) &= \frac{1}{\sqrt{2(\psi-\mathcal{E})}} \int_0^\mathcal{E} \mathrm{d}\mathcal{E'} \left(\frac{\psi-\mathcal{E'}}{\psi-\mathcal{E}}\right)^{(k-1)/2} f_\alpha (\mathcal{E'}) \, .
    \end{align}
    Here $f$ depends only on $\mathcal{E}$ as we assume that the distribution function is ergodic, meaning it depends on $r$ and $v$ only through the Hamiltonian: $f(r,v)=f \big( H(r,v) \big)$.

    The ergodic distribution function corresponding to a spherical density profile can be computed using Eddington's formula \citep{1916MNRAS..76..572E, 2013degn.book.....M}:
    \begin{equation}
        f_\alpha(\mathcal{E}) = \frac{1}{\sqrt{8}\pi^2} \left(\frac{1}{\sqrt{\mathcal{E}}} \left.\frac{\mathrm{d}n_\alpha}{\mathrm{d}\psi}\right\rvert_{\psi = 0} + \int^\mathcal{E}_0 \frac{\mathrm{d}\psi}{\sqrt{\mathcal{E}-\psi}}\frac{\mathrm{d}^2 n_\alpha}{\mathrm{d}\psi^2}\right) \, .
    \end{equation}
    We can simplify this expression by observing that $\psi = 0$ when $r \to \infty$. For a Dehnen density profile, at infinite distance we have that $n_\alpha \propto \psi^4$, since $n_\alpha \propto r^{-4}$ and $\psi \propto r^{-1}$. Thus,
    \begin{equation}
        \left.\frac{\mathrm{d}n_\alpha}{\mathrm{d}\psi}\right\rvert_{\psi=0} = \left.4 \psi^3\right\rvert_{\psi=0} = 0 \, ,
    \end{equation}
    which gives us the simpler expression:
    \begin{equation}
        f_\alpha(\mathcal{E}) = \frac{1}{\sqrt{8}\pi^2} \int^\mathcal{E}_0 \mathrm{d}\psi \, g(\psi) \, ,
    \end{equation}
    where
    \begin{equation}
        g(\psi) = \frac{1}{\sqrt{\mathcal{E}-\psi}}\frac{\mathrm{d}^2 n_\alpha}{\mathrm{d}\psi^2} \, .
    \end{equation}
    Finally, we can observe that:
    \begin{equation}
        \frac{\mathrm{d}^2 n_\alpha}{\mathrm{d}\psi^2} = \frac{\mathrm{d}r}{\mathrm{d}\psi} \frac{\mathrm{d}}{\mathrm{d}r} \left(\frac{\mathrm{d}r}{\mathrm{d}\psi} \frac{\mathrm{d}}{\mathrm{d}r} n_\alpha\right) = \frac{1}{\left(\psi '\right)^2} n''_\alpha-\frac{\psi''}{\left(\psi'\right)^3}n'_\alpha \, ,
    \end{equation}
    where the prime symbol indicates derivation with respect to $r$.

    In practice, to find $f_\alpha$ we have to integrate $g(\psi)$ over $\psi$ between $0$ and $\mathcal{E}$. We perform this numerically by evaluating the function $g(\psi)$ at discrete values $\psi_i$ inside the range $[0, \mathcal{E}]$. Thus, to use the expression for the second derivative of $n_\alpha$ that we just found, for each of these values we first numerically find the $r_i$ such that $\psi(r_i) = \psi_i$ by solving Eq.\ \eqref{eq:psi}. Then, we can compute $\psi'(r_i)$, $\psi''(r_i)$, $n'(r_i)$, and $n''(r_i)$, as their analytic expressions can be easily obtained from Eqs.\ \eqref{eq:n}, \eqref{eq:phi}, and \eqref{eq:psi}.

    In order to estimate the orbit-averaged relaxation timescale, we compute the orbit averaged diffusion coefficient $\mathrm{D}_\mathrm{o}[\Delta \mathrm{R}]$ as
    \begin{equation}
        \mathrm{D}_\mathrm{o}[\Delta \mathrm{R}] = \lim_{R\to 0} \frac{D_{RR}(\mathcal{E}, R)}{4\pi^2\,J_\mathrm{c}^2(\mathcal{E})\,P(\mathcal{E}, R)} \, ,
    \end{equation}
    where $D_{RR}$ is one of the coefficients of the orbit-averaged Fokker-Planck equation in the flux conservation form. It can be computed as
    \begin{equation}
    \begin{split}
        D_{RR} &= \frac{256}{3}\, \pi^4 \, G^2\, \log \Lambda\,R\,J^2 \,  \times \\
        &\quad \sum_{\alpha} m_\alpha^2 \int_{r_-}^{r_+} \frac{v\, \mathrm{d}r}{v_\mathrm{r}} \left[\left( \frac{2}{v_\mathrm{t}^2} + \frac{2\,v^2}{v_\mathrm{c}^4}- \frac{4}{v_\mathrm{c}^2} \right) \mathcal{J}_{1,\alpha}+ \left( \frac{3}{v_\mathrm{t}^2} - \frac{3}{v^2} \right) \mathcal{I}_{2,\alpha} \right.\\
        &\qquad \qquad + \left( \frac{3}{v^2} - \frac{1}{v_\mathrm{t}^2} \left. + \frac{2\,v^2}{v_\mathrm{c}^4} - \frac{4}{v_\mathrm{c}^2}  \right) \mathcal{I}_{4,\alpha}\right] \, ,
    \end{split}
    \end{equation}
    where $v_\mathrm{c} = J_\mathrm{c} / r$ is the circular velocity at a given energy.

\section{From $\delta v_\parallel$ and $\delta v_\perp$ to $\delta \mathcal{E}$ and $\delta J$}\label{app:VtoEJ}

    The aim of this appendix is to show how we can express the change in energy and angular momentum of the orbiting stellar-mass BH as a function of its velocity change in the parallel and perpendicular directions relative to its original motion.

    At first, the velocity of the stellar-mass BH is
    \begin{equation}
        \vec{v} = v_\mathrm{r} \vec{\hat{e}_\mathrm{r}} + v_\mathrm{t} \vec{\hat{e}_\mathrm{t}} \, ,
    \end{equation}
    where $\vec{\hat{e}_\mathrm{r}}$ and $\vec{\hat{e}_\mathrm{t}}$ respectfully represent the radial and tangential unit vectors of the orbit.

    The variation of velocity $\delta \vec{v}$ due to a stochastic velocity kick is
    \begin{equation}
        \delta \vec{v} = \left( \delta v_\parallel \frac{v_\mathrm{r}}{v} - \delta v_\perp \sin \theta \frac{v_\mathrm{t}}{v} \right) \vec{\hat{e}_\mathrm{r}} + \left( \delta v_\parallel \frac{v_\mathrm{t}}{v} + \delta v_\perp \sin \theta \frac{v_\mathrm{r}}{v} \right) \vec{\hat{e}_\mathrm{t}} + \delta v_\perp \cos \theta \vec{\hat{e}_\mathrm{3}} \, ,
    \end{equation}
    where $\vec{\hat{e}_\mathrm{3}}$ is the unit vector in the direction perpendicular to the orbital plane and $\theta$ is the angle between a direction perpendicular to the motion of the stellar-mass BH and $\vec{\hat{e}_\mathrm{3}}$. In practice, in our implementation this angle is always uniformly selected in the range $[0,2\pi)$, thus we do not properly define the direction of $\delta v_\perp$ here.

    We can now compute the change in energy given $\delta v_\parallel$ and $\delta v_\perp$:
    \begin{equation}
        \begin{aligned}
            \delta \mathcal{E} &= -\frac{(\vec{v}+\delta \vec{v})^2}{2} + \frac{\vec{v}^2}{2} = -\frac{(\delta \vec{v})^2}{2} + \vec{v} \cdot \delta \vec{v} \\
            &= -v \delta v_\parallel - \frac{1}{2} (\delta v_\parallel)^2 - \frac{1}{2} (\delta v_\perp)^2 \, .
        \end{aligned}
    \end{equation}
    The vector change of $\vec{J}$ is instead
    \begin{equation}
        \delta \vec{J} = r \vec{\hat{e}_\mathrm{r}} \times \delta \vec{v} = - r \delta v_\perp \sin \theta \vec{\hat{e}_\mathrm{t}} + r \left(\frac{v_\mathrm{t}}{v} \delta v_\parallel + \frac{v_\mathrm{r}}{v} \delta v_\perp \cos \theta \right) \vec{\hat{e}_\mathrm{3}} \, ,
    \end{equation}
    from which follows that $J=|\vec{J}|$ changes as
    \begin{equation}
        \begin{aligned}
            \delta J &= |\vec{J} + \delta \vec{J}| - |\vec{J}| \\
            &= J \left( \frac{\delta v_\parallel}{v} + \delta v_\perp \frac{v_\mathrm{r}}{v v_\mathrm{t}} \sin{\theta} +\frac{1}{2} \left(\frac{\delta v_\perp \cos{\theta}}{v_\mathrm{t}}\right)^2 \right) + \mathcal{O} \left( (\delta v)^3 \right) \, ,
        \end{aligned}
    \end{equation}
    where we used $J=rv_\mathrm{t}$.

    More detailed steps can be found in \citet{broggi_phd}.\footnote{The expression for $\delta J$ reported in \citet{broggi_phd} mistakenly reports an extra term proportional to $\delta v_\parallel \delta v_\perp$.}

\section{From $\Delta \mathcal{E}$ and $\Delta J$ to $\Delta v$}\label{app:EJtoV}
    The aim of this appendix is to show how we can kick the stellar-mass BH to recover a given change in energy and angular momentum. Crucially, we wish to achieve this only by changing the velocity of the stellar-mass BH, while keeping its position the same.

    At first, the velocity of the stellar-mass BH is
    \begin{equation}
        \vec{v} = v_\mathrm{r} \vec{\hat{e}_\mathrm{r}} + v_\mathrm{t} \vec{\hat{e}_\mathrm{t}} \, ,
    \end{equation}
    where $\vec{\hat{e}_\mathrm{r}}$ and $\vec{\hat{e}_\mathrm{t}}$ respectfully represent the radial and tangential unit vectors of the orbit. The most generic velocity kick we can give to the stellar-mass BH is
    \begin{equation}
        \Delta \vec{v} = \Delta v_\mathrm{r} \vec{\hat{e}_\mathrm{r}} + \Delta v_\mathrm{t} \vec{\hat{e}_\mathrm{t}} +  \Delta v_\mathrm{3} \vec{\hat{e}_\mathrm{3}}\, ,
    \end{equation}
    where $\vec{\hat{e}_\mathrm{3}}$ is the unit vector in the direction perpendicular to the orbital plane. Thus, the new velocity of the stellar-mass BH $\vec{v_1}$ is
    \begin{equation}
        \vec{v_1} = ( v_\mathrm{r} + \Delta v_\mathrm{r} ) \vec{\hat{e}_\mathrm{r}} + ( v_\mathrm{t} + \Delta v_\mathrm{t} ) \vec{\hat{e}_\mathrm{t}} +  \Delta v_\mathrm{3} \vec{\hat{e}_\mathrm{3}} \, .
    \end{equation}
    In Newtonian dynamics, $\vec{J} = \vec{r} \times \vec{v}$. Thus,
    \begin{equation}
        \vec{J} = r v_\mathrm{t} \vec{\hat{e}_\mathrm{3}}
    \end{equation}
    and
    \begin{equation}
        \vec{J_1} = \vec{r} \times \vec{v_1} = -r \Delta v_\mathrm{3} \vec{\hat{e}_\mathrm{t}} + r ( v_\mathrm{t} + \Delta v_\mathrm{t} ) \vec{\hat{e}_\mathrm{3}} \, ,
    \end{equation}
    where $\vec{J_1}$ represents the angular momentum after the change in velocity. Moreover, the energy will change from
    \begin{equation}
        \mathcal{E} = - \frac{v^2}{2} + \psi(r) 
    \end{equation}
    to
    \begin{equation}
        \mathcal{E}_1 = - \frac{v_1^2}{2} + \psi (r) \, .
    \end{equation}
    
    Now, we write the following system:
    \begin{equation}
        \begin{cases}
            \displaystyle \mathcal{E}_1 = \mathcal{E} + \Delta \mathcal{E} \\
            \displaystyle J_1 = J + \Delta J \\
            \displaystyle \frac{\vec{J_1} \cdot \vec{J}}{J_1 J} = \cos{\beta}
        \end{cases} \, ,
    \end{equation}
    where $\beta$ is the angle between vectors $\vec{J_1}$ and $\vec{J}$. Using the relations found above, we can rewrite the system as
    \begin{equation}
        \begin{cases}
            \displaystyle -\frac{1}{2} \sqrt{(v_\mathrm{r}+\Delta v_\mathrm{r})^2 + (v_\mathrm{t}+\Delta v_\mathrm{t})^2 + (\Delta v_\mathrm{3})^2} = -\frac{1}{2} v^2 + \Delta \mathcal{E} \\
            \displaystyle r \sqrt{(\Delta v_\mathrm{3})^2 + (v_\mathrm{t} + \Delta v_\mathrm{t})^2} = J + \Delta J \\
            \displaystyle \frac{r (v_\mathrm{t} + \Delta v_\mathrm{t}) }{J + \Delta J} = \cos{\beta}
        \end{cases} .
    \end{equation}
    In order to solve this system, we need to know the value of $\beta$. In practice, during our simulations we randomly drew it from the interval $[0, 2 \pi)$, thus assuming an isotropic distribution.
    
    Finally, after a sequence of substitutions, the system above can be written as
    \begin{equation}
        \begin{cases}
            \displaystyle \Delta v_\mathrm{r} = \sqrt{v^2 - 2 \Delta \mathcal{E} - \left(\frac{J + \Delta J}{r}\right)^2} - v_\mathrm{r} \\
            \displaystyle \Delta v_\mathrm{t} = \frac{(J + \Delta J) \cos{\beta}}{r} - v_\mathrm{t} \\
            \displaystyle \Delta v_\mathrm{3} = \frac{(J + \Delta J) \sin{\beta}}{r}
        \end{cases} \, .
    \end{equation}

\end{appendix}

\end{document}